\documentclass[twocolumn]{aastex7}
\usepackage{graphicx}
\usepackage{subcaption}
\usepackage{natbib}
\usepackage{hyperref}
\usepackage{amsmath}
\usepackage[T1]{fontenc}
\usepackage[utf8]{inputenc}

\newcommand{\msun}{$M_{\sun}$}
\newcommand{\rsun}{R$_{\sun}$}
\newcommand{\disk}{_{\text{disk}}}
\newcommand{\ejec}{_{\text{ejec}}}

\newcommand{\WD}{_{\text{WD}}}
\newcommand{\iso}[2]{$^{#2}$\text{#1}}

\graphicspath{{./}{figures/}}

\received{April 21, 2025}
\revised{July 31, 2025}
\accepted{August 11, 2025}

\shorttitle{The interaction between the nova/supernova ejecta and the accretion disk}
\shortauthors{Lechuga et al.}

\begin{document}

\title{High-Resolution Simulations of the Interaction Between the Nova/Supernova Ejecta and the Accretion Disk}

\author[0009-0006-7500-2723]{Axel S. Lechuga}
\affil{Departament de Física. Universitat Politècnica de Catalunya (UPC) \\
 Av. Eduard Maristany 16, 08019 Barcelona, Spain}
\email[show]{axel.sanz@upc.edu}

\author[0000-0001-5197-7100]{Domingo García-Senz}
\affil{Departament de Física. Universitat Politècnica de Catalunya (UPC) \\
 Av. Eduard Maristany 16, 08019 Barcelona, Spain}
 \affil{Institut d’Estudis Espacials de Catalunya (IEEC) \\ 08860 Castelldefels (Barcelona), Spain}
\email[show]{domingo.garcia@upc.edu}

\author[0000-0002-9937-2685]{Jordi José}
\affil{Departament de Física. Universitat Politècnica de Catalunya (UPC) \\
 Av. Eduard Maristany 16, 08019 Barcelona, Spain}
  \affil{Institut d’Estudis Espacials de Catalunya (IEEC) \\ 08860 Castelldefels (Barcelona), Spain}
  \email[show]{jordi.jose@upc.edu}

\begin{abstract}

Many nova and type Ia supernova explosion scenarios involve accretion disks. However, direct numerical simulations of these explosive phenomena have barely addressed the question of the impact of ejecta-disk collision on the midterm evolution of such explosions. This is particularly critical for a better understanding of classical and recurrent novae, where each nova cycle depends on the imprint left by the precedent explosion. In this work, we describe and analyze a set of high-resolution simulations of the ejecta-disk interaction. We show that, depending on the initial configuration of the binary system, the disk is partially or, more often, totally destroyed, which will impact the next nova-explosion cycle. In the case of type Ia supernovae, the much larger kinetic energy carried by the ejecta always provokes complete destruction of the accretion disk. We also discuss the alterations induced in the geometry of the ejecta by the shielding effect of the disk, which has shown to cause reduced contamination of the companion star up to a factor $\sim 1.5-2$ in key nuclei produced during the nova outburst. In the framework of recurrent nova simulations, we report for the first time on the formation of a cavity in the ejecta after its interaction with the disk. We also describe the onset and development of several hydrodynamic instabilities such as Kelvin-Helmholtz and Richtmyer-Meshkov.
  
\end{abstract}


\section{Introduction} 

Classical novae (CNe) are explosive phenomena that take place on the surface of white dwarfs (WD) that accrete hydrogen-rich material from a non-degenerate main-sequence companion star, via an accretion disk formed when angular momentum-rich matter from the secondary star overflows the Roche-Lobe \citep[see][for reviews]{starrfield2008,starrfield2016,jose2008,jose2016,chomiuk2021}. These binary systems typically have short orbital periods that range from approximately 1 to 15 hours. When enough material accumulates on the surface of the WD, the temperature and pressure at the base of the accreted envelope are such that nuclear reactions are triggered (initially by pp chains, followed by CNO cycles). Due to the mildly degenerate conditions in the accreted envelope, the material continues to increase temperature without expanding, setting in nuclear fusion, and driving a thermonuclear runaway (TNR) that leads to the final nova outburst \citep{starrfield1972, prialnik1995, jose1998}. The outburst is strong enough to eject the outer $10^{-7}-10^{-4}$ M$_\sun$ with velocities of the order of 1000 km s$^{-1}$, powering an event that reaches peak luminosities of $10^4-10^5$ $L_{\sun}$. 

Following the eruption, the WD remains mostly intact, having ejected only a fraction of its outermost layers. As accretion from the companion star resumes, the WD can gradually rebuild its hydrogen envelope and begin a new nova cycle with typical recurrence periods of $10^4 - 10^5$ years, according to current nova models. However, a subclass of novae known as recurrent novae (RNe), by definition\footnote{More recently, another definition based on the presence of vast supershells ejected in previous outbursts has also been proposed \citep{Pagnotta2014,Schaefer2022}.} novae observed in outburst more than once, have much shorter recurrence periods between $~1-100$ years. RNe are symbiotic systems consisting of an accreting WD and a red giant companion. About 10 RNe have been discovered in the Milky Way, although several others have been observed in the Andromeda Galaxy and the Large Magellanic Cloud \citep{Darnley_Henze2020}. RNe exhibit a much wider range of orbital periods compared to CNe, spanning from a few hours to several hundred days. Various classifications have been proposed, based on different orbital periods, the presence of a plateau in the tail of the light curve, varying amplitudes and recurrence times, or distinct mechanisms driving the high mass-accretion rates inferred \citep[see][and references therein]{schaefer2010}. The mass transfer in RNe proceeds either through Roche-Lobe overflow or stellar wind from the evolved secondary star, as a result not all RNe have an accretion disk.

Accretion onto a WD can also lead to a type Ia supernova (SNIa).
This class of supernovae is characterized by the absence of hydrogen in the spectra \citep{1990RPPh...53.1467W,1997ARA_A..35..309F,annurev:/content/journals/10.1146/annurev.astro.36.1.17}, which puts constraints on the maximum amount of this element that can be present in the expanding atmosphere roughly ${\rm M_H} \leq$ 0.03 - 0.1 M$_\odot$. Some SNIa (e.g., SN 2002ic), however, reveal H emission lines some months after achieving maximum brightness, whose origin has been attributed to interactions between the supernova ejecta and H-rich circumstellar material \citep{silverman2013}. SNIa also exhibit a distinctive absorption feature near 6150 \AA, due to blueshifted Si II, absent in other type I subclasses.
Two scenarios have been proposed to explain the origin of SNIa: a {\it single-degenerate channel} (SD), involving transfer of H- or He-rich matter from a low-mass star onto a CO WD, and a {\it double-degenerate channel} (DD), where two CO WDs merge as a result of energy and angular momentum losses driven by the emission of gravitational waves \citep[see][for a recent review]{ruiter2025}. The progenitor in the SD scenario bears some similarities with that of current novae because the ejected material first impacts the accretion disk and subsequently with the non-degenerate companion star, which is the one considered in this work. However, the energy released in a SNIa, approximately ${\rm E_{kin}} \sim 10^{51}$ erg, with expansion velocities typically between 5000 and 10000 km s$^{-1}$, and the mass involved in the ejected plasma (i.e., the entire WD) far exceed the values that characterize CNe and RNe.

During a nova outburst, the disk is susceptible to be altered or even destroyed due to the impact of the ejecta. Thus, the degree of disk disruption and its ability to recover will impact how mass transfer resumes and how the system continues its nova cycle. Along with the accretion rate, the survivability of the disk directly influences the recurrence time of nova events; in systems where it is significantly disrupted, the recovery time to rebuild the disk increases, resulting in longer periods. Several observations report signatures of disk disruption and reformation following nova explosions. For example, the RN system U Sco has shown mass transfer recovery 8-10 days after the outburst with narrow emission lines arising from reforming disk material \citep{Mason2012,Schaefer_etal2011}. Similar disk reformation signatures have been identified in systems like CI Aql, T Pyx, and RS Oph \citep{Schaefer_2011,Tofflemire_etal2013,Azzollini2023}.

Another critical aspect of the evolution of the accretion disk after a nova outburst is its impact on the angular momentum of the system. During expansion, the ejecta will collide with the disk and transfer momentum that can affect the disk rotational dynamics, potentially causing sections of the disk to move outwards. This redistribution can destabilize the disk internal structure and cause different phenomena (hydrodynamic instabilities, turbulence, transient disk geometries), resulting in a potential acceleration or inhibition of mass accretion with consequences for future nova cycles. Moreover, these altered accretion states during the recovery and stabilization phases of the disk can have broader implications for the long-term evolution of the binary system. During the next cycle, irregular accretion might impact the development of the next TNR which can lead to abnormal outbursts, likely causing a mass imbalance between the accreted and ejected material. It is widely accepted that mixing plays a key role in determining the fate of an accreting WD during nova outbursts. In CNe, significant mixing at the WD-envelope interface leads to the ejection of more material than it is accreted, ultimately reducing the WD mass. In contrast, RNe experience limited mixing, which minimizes mass ejection and allows the WD to grow in mass; as a result, the WD may eventually reach the Chandrasekhar limit, making RNe likely type Ia supernova progenitors \citep{Soraisam2015,Jose_Hernanz2025}.

Observational evidence supports the idea that the
nova ejecta interacts with the surrounding environment,
such as the gamma-ray emission detected at energies exceeding 100 MeV.
In symbiotic systems, such as V407 Cyg or RS OPh, this high-energy emission has been attributed to shock acceleration in the ejected shells
after interaction with the dense wind of the red giant companion
\citep{HernanzTatischeff+2012+62+67}. The emission reported from several CNe involving less evolved stellar companions (e.g., V1324 Sco, V959 Mon, V339 Del, and V1369 Cen; \citealt{Ackerman2013}) has been attributed to internal shocks in the ejecta. The nova ejecta is expected to interact with the accretion disk, the companion star and the circumstellar matter. Some systems, such as RS Oph, show asymmetric ejection patterns and
bipolar structures in their nova remnant \citep{montez2022}, suggesting a strong interaction with the environment
and where the disk may be involved in shaping
the ejecta and determining its final geometry. Additionally,
by absorbing or deflecting some of the ejected mass,
the accretion disk can reduce the deposition of intermediate-mass
elements from the nucleosynthetic yields produced in
the outburst onto the companion star, i.e. act as a
shield that mitigates the chemical pollution caused by
the metal-rich material into the atmosphere of the companion star,
and hence impact its spectral characteristics.

Pioneering numerical simulations analyzing the role of the accretion disk during the outburst in CNe \citep{figueira2018} and RNe \citep{figueira2025} have highlighted the importance of including the accretion disk to describe the long-term phase of the explosion. Those simulations were performed in three dimensions (3D) and, therefore, free of geometrical restrictions. However, they are heavily expensive in terms of computational cost and often very limited in resolution, which is typically not enough to capture the rich phenomenology that takes place in the interaction.
In order to better understand the complex dynamics between the nova ejecta and the accretion disk, it is necessary to perform high-resolution simulations that can capture the fine-scale interactions and structural changes occurring during the post-nova stage. To achieve this, an axisymmetric approach has been used in this work, which allows for higher spatial resolution while maintaining a sufficient level of physical consistency. 

Section \ref{sec:modelling} is devoted to the models and input physics used in this work to study the interaction between the ejecta and the accretion disk. Section \ref{sec:numericalapproach} is dedicated to the numerical approach and the simulation methods.  Section \ref{sec:results} contains all the results, focusing on the evolution of the collision, the survivability of the disk, the impact on the geometry of the ejecta and the pollution effect on the secondary star. A summary of the main conclusions is given in section \ref{sec:conclusions}. Additional details regarding the building of the intial models are given in the appendix.

\section{Model and Input Physics }
\label{sec:modelling}

\subsection{Initial Models}

\begin{figure}
    \centering
\includegraphics[width=1.0\linewidth]{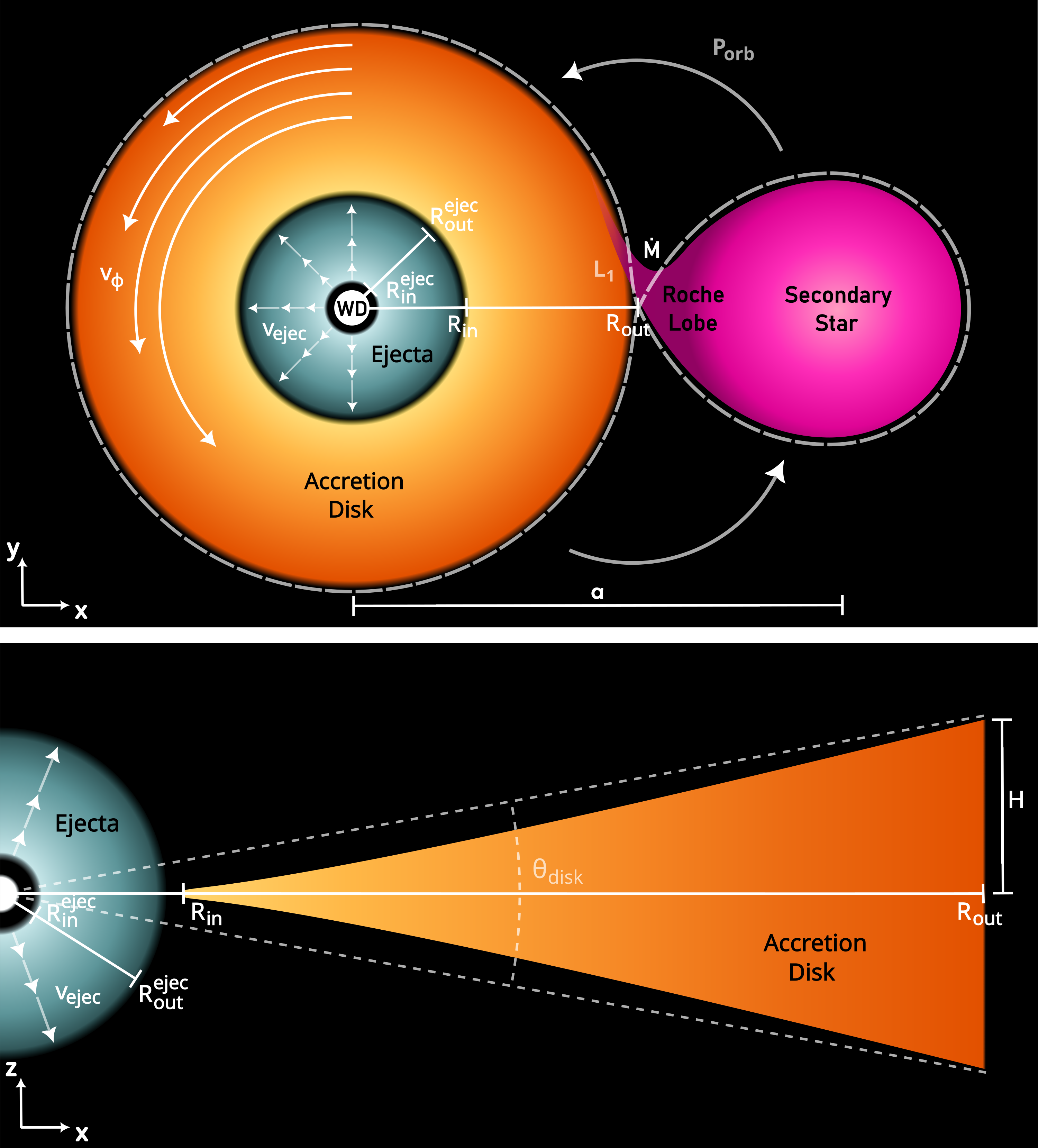}
    \caption{Orbital and axial view of the binary systems (not in scale). Only the ejecta and the accretion disk are included in the simulations. The values adopted for the different magnitudes are given in Tables \ref{tab:parameters1} \& Table \ref{tab:parameters2} }
    \label{fig:scheme}
\end{figure}

An overview of the scenario of our calculations is shown in Fig.~\ref{fig:scheme}, where the top panel shows the more relevant elements in the orbital plane. These include the WD, the nova/supernova ejecta, the accretion disk and the companion star. The bottom panel shows an axial view, although restricted to the nova/supernova ejecta and the disk, which are the objects considered in the axisymmetric simulations presented in this work. We have carried out a set of simulations of a handful of representative cases, including two recurrent novae and four classical novae. The simulations are built taking into account plausible combinations of the relevant parameters describing the WD-ejecta-disk system with the aim of studying their impact in the medium-term evolution of the outburst. A summary of the models computed can be found in Tables \ref{tab:parameters1} and \ref{tab:parameters2}.

\begin{deluxetable*}{c|ccccccccc}[t!]
\tablecaption{Binary system modelling parameters\label{tab:parameters1}}
\tablehead{\colhead{Name} & \colhead{$\text{WD}_\text{comp}$} & \colhead{$M\WD$} &\colhead{$P_{orb}$}  & \colhead{$a$} & \colhead{$\dot{M}$} & \colhead{$M\disk$} & $R_{in}$ & $R_{out}$ &\colhead{$\theta\disk$} \cr ... & ... &  [\msun] & [days]/[h] & [\rsun]& [\msun\, yr$^{-1}$] & [\msun] & [\rsun] & [\rsun] & [$^\circ$]}
\startdata
        RN1 & O-Ne & 1.38 & 454 days& 332.83 &2E-7 & 4.84E-6 & 16.64 & 190.75  &  13.17\\
        RN2 & O-Ne & 1.38 & 5 days & 16.43 &2E-7& 1.13E-7 & 0.82 & 9.41  &   9.05\\
        CN1 & C-O & 0.6 & 10 h &
        2.65 &2E-10 & 4.18E-11 & 0.13 & 1.05 & 3.30\\
        CN2 & C-O  & 1.25 & 10 h  &
        3.08 &2E-10 & 1.03E-10 & 0.15 &1.69 &  2.70\\
        CN3 & O-Ne  & 1.25 & 10 h  &
        3.08 &2E-10 & 1.03E-10 & 0.15 & 1.69 &  2.70\\
        CN4 & O-Ne  & 1.25 & 10 h & 3.08 &2E-10 & 1.03E-10 & 0.15 & 1.69 &  2.70\\
        SNIa & C-O & 1.36 & 7.78 h & 2.65 &5E-7 & 1.42E-8 & 0.14 & 1.52 &  7.86\\  
\enddata
\tablecomments{$\text{WD}_\text{comp}$ and $M\WD$ denote the chemical composition and mass of the WD. $P_{orb}$ and $a$ are the orbital period and separation of the binary system, considering for all models a 1 \msun $ $ companion star. $\dot{M}$ is the accretion rate onto the WD, we have chosen typical values in CNe, RNe and SNIa models. Lastly $M\disk$, $R_{in}$, $R_{out}$ and $\theta\disk$ are the mass, position of the inner and outer edges, and aperture angle of the accretion disk, respectively.}
\end{deluxetable*}

The first two models RN1 and RN2 are examples of RNe. The former resembles the system RS Ophiuchi, formed by a WD and a red giant companion of approximately 106 \rsun $ $ \citep{Barry2008} and orbital period $P_{orb} = 454$ days. The latter is a distinctive RNe system with $P_{orb} = 5$ days. The difference between these two models is only the orbital period, spanning the wide range of values in known RNe systems. This translates into different distances between both stars: $2.32\times{10}^{13}$ cm (332.83 \rsun) and $1.14\times{10}^{12}$  cm (16.43 \rsun) respectively.

The other simulations correspond to CNe, in which the ejected mass is orders of magnitude higher (at lower ejection velocities) than in typical RNe. Model CN1 resembles a case of a $0.6$ \msun $ $ C-O WD. Model CN2 is identical to CN1 except for the considerably larger mass of the WD, $1.25$ \msun \footnote{Note that this value its too large for a C-O WD but this model is aimed at checking the effect of the WD mass alone in the simulation.}. The adopted masses are representative for CO and ONe WDs and lay within the range of masses inferred in novae. Since the orbital period is the same, this results in a slightly larger system that changes the structure of the accretion disk which becomes larger and denser.

The model CN3 is analogous to model CN2 but with significantly less ejected mass (and similar ejecta velocities), which is useful to test the effect of the total mass impacting the disk. The last model is CN4, which is a typical example of an O-Ne nova and is used to check the impact of having higher ejecting velocities when compared to model CN3.

We also present a calculation involving a type Ia Supernova. In this case, it is expected that the much larger mass, velocity and kinetic energy of the ejecta promises complete destruction of the accretion disk, which is confirmed by our simulations.

\subsection{Ejecta}

The different ejecta have been modelled following the density, velocity, temperature and chemical composition profiles obtained from the hydrodynamic code {\tt SHIVA}, a hydrodynamic, Lagrangian, finite-difference, time-implicit code \citep[see][for details]{jose1998,jose2016}. {\tt SHIVA} solves the standard differential equations of stellar evolution---mass, energy, and momentum conservation, as well as energy transport---and has been extensively used for over 25 years in studies of stellar explosions, such as CNe and RNe, type I X-ray bursts, and (sub-Chandrasekhar) SNIa. The code uses a general equation of state that includes contributions from the degenerate electron gas, multicomponent ion plasma, and radiation \citep{Blinnikov1996}. It incorporates Coulomb corrections to electron pressure. Nuclear energy generation is modeled using a reaction network of 120 isotopes (ranging from $^1$H to $^{48}$Ti), linked by 630 nuclear processes, with updated reaction rates from the STARLIB database \citep[][Iliadis, priv. comm.]{Sallaska2013}. Screening factors and neutrino energy losses are also included. {\tt SHIVA} also incorporates a time-dependent convective transport formalism that sets in when the characteristic convective timescale exceeds the integration timestep. Partial mixing between adjacent convective shells is modeled using a diffusion equation \citep{Prialnik1979}. The code is able to track a full nova cycle from accretion, ignition, expansion and final ejection. In all calculated cases we set the outermost region of the ejecta almost in contact with the inner radius of the accretion disk so that the collision begins shortly after $t = 0$ s.

\begin{deluxetable}{c|cccccc}[t]
\tablecaption{Ejecta modelling parameters\label{tab:parameters2}}
\tablehead{\colhead{Name} & \colhead{$M\ejec$} & \colhead{$\tilde{v}\ejec$} &\colhead{$\tilde{\mu}\ejec$ } & \colhead{$E_{kin}$} & \colhead{$p$} & $\tau$ \cr ... & [\msun]  &[km s$^{-1}$]& ... & [erg] & [g cm s$^{-1}$] & [min]}
\startdata
        RN1 & 1.18E-6 & 4180 & 0.674 & 2.05E44 & 9.82E35 & 529.8\\
        RN2 & 1.18E-6 & 4180 & 0.674 & 2.05E44 & 9.82E35 & 26.13\\
        CN1 & 2.08E-4 & 670 & 0.928 & 2.02E45 & 2.79E37 & 18.19\\
        CN2 & 2.08E-4 & 670 & 0.928 & 2.02E45 & 2.79E37 & 29.27\\
        CN3 & 2.06E-5 & 820 & 0.997 & 1.67E45  & 3.34E36 & 23.91 \\
        CN4 & 2.06E-5 & 2580 & 0.997 & 1.67E44 & 1.056E37 & 7.60\\
        SNIa & 1.36 & 7100 & 40.06 &  8.05E50 & 1.97E42 & 2.48\\  
\enddata
\tablecomments{Columns correspond to ejected mass, mass-averaged velocity of the ejecta, mean molecular weight, kinetic energy, radial momentum and characteristic time. Notice the different energies and radial momentum of the ejecta due to the changes in ejected mass and velocities. $\tau = R_{\text{out}}/\tilde{v}\ejec$ is taken as a characteristic time for the collision, used for computing the dimensionless time $t'=t/{\tau}$ for a better comparison between scenarios. The values were obtained with the {\tt SHIVA} code.}
\end{deluxetable}

\begin{figure*}[h!]
    \centering 
    \includegraphics[width=1\linewidth]{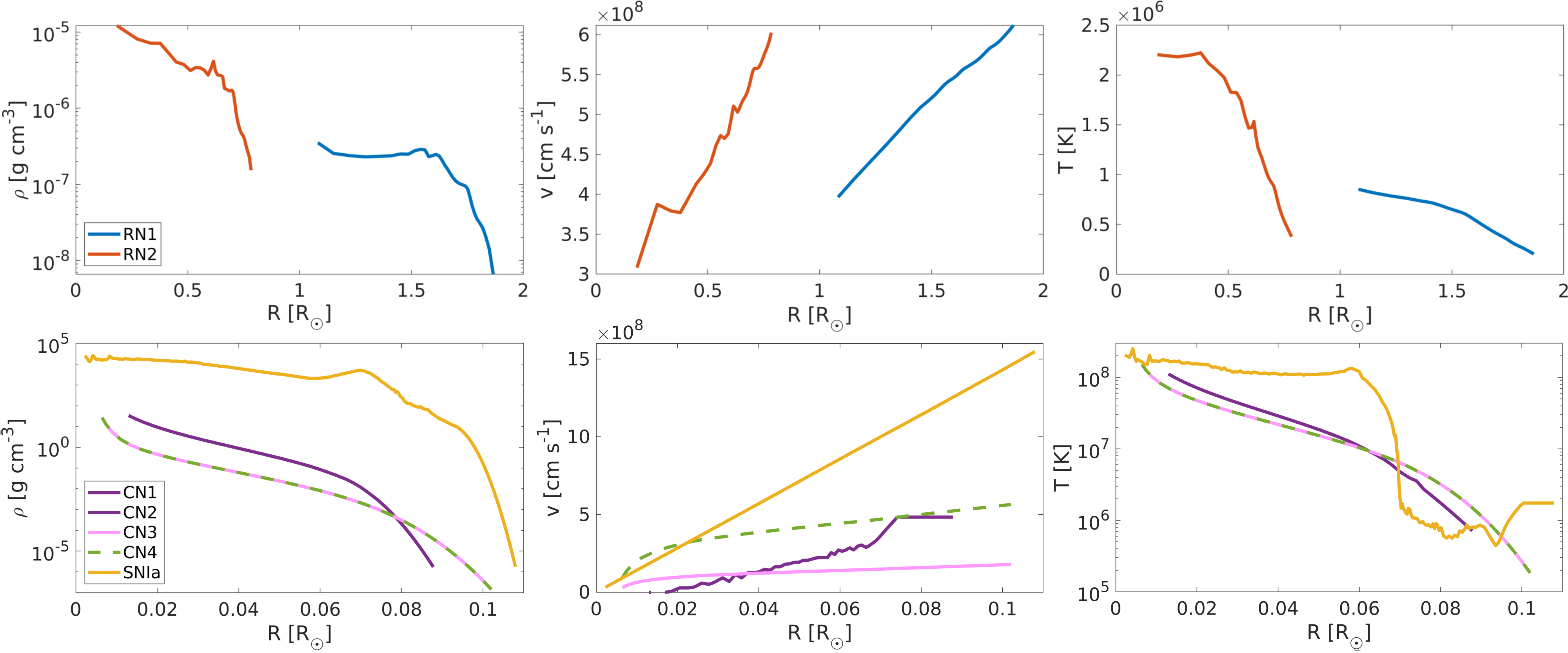}
    \caption{Hydrodynamic profiles of the ejecta in the different models computed in this work. Models CN1 and CN2 share the same ejecta and differ only in the properties of their accretion disks.}
    
    \label{fig:profiles_ejecta}
\end{figure*}

Figure \ref{fig:profiles_ejecta} shows the density, velocity and temperature profiles of the five different ejecta types considered in this work. The location of the outermost part of the ejecta is chosen depending on the position of the inner edge of the disk $R_{\text{in}}$. This can be seen in the top RNe row where the same ejecta (same parameters) is taken at different expansion times. It is worth noting that, at the moment of the collision, the different velocity profiles have reached, or are close to, the homologous phase, as expected from point-like explosions after sufficiently long elapsed times.

    \begin{figure*}[h!]
    \centering
    \includegraphics[width=1\linewidth]{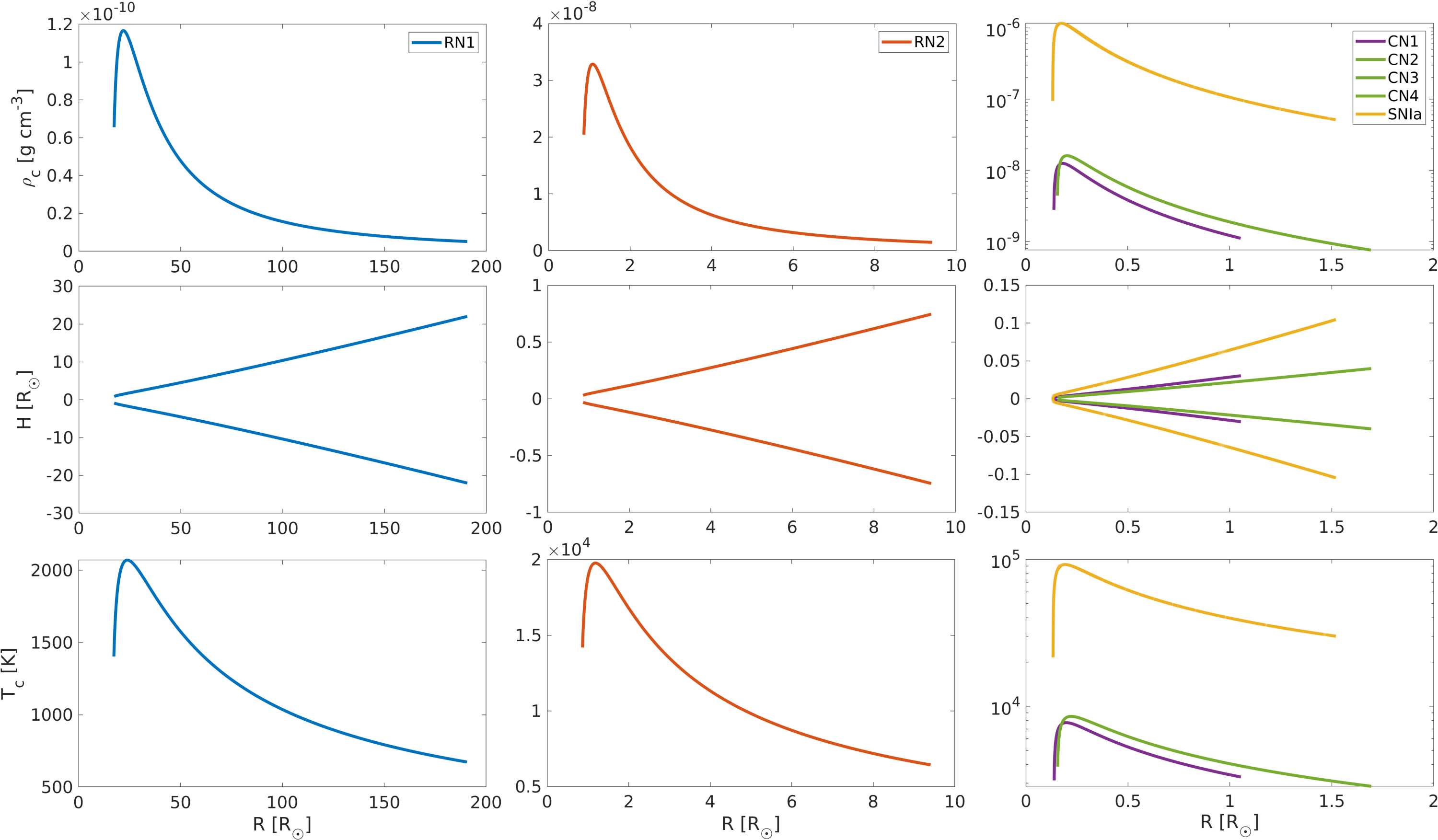}
    \caption{Density, height and temperature profiles of the different accretion disks considered in this work. The subscript in $\rho_c$ and $T_c$ refers to the value of the magnitude at the center (z=0). Notice the different sizes in the x-axis due to the different orbital periods considered, as well as the effect of the high accretion rate in the SNIa case.}
    \label{fig:profiles_disk}
\end{figure*}

\subsection{Accretion Disk}

The accretion disk has been modelled following the Shakura-Sunyaev thin disk solution \citep{Shakura1973}, which consists in a thin disk (height $H << $ radius R),  with negligible self-gravity ($M\disk <<  M\WD$). This model is based on the well-known $\alpha$-prescription, where the viscous effects caused by unresolved turbulent behavior in the angular momentum transfer mechanism are parametrized as $\tau_\nu=\alpha\rho c_s^2$, with $\tau_\nu$ being the viscosity of the plasma, $\rho$ the density, $c_s$ the speed of sound and $\alpha$ a parameter that ranges from 0 to 1. Although typical values of $\alpha$ in cold disk modeling remain somewhat uncertain, observational and theoretical studies suggest values ranging from $\sim0.1-0.4$ \citep{King2007,Kotko2012}. For our simulations, we adopt a moderate value of $\alpha = 0.3$, which is well within this established range.

The outer radius of the disk $R_{out}$ is considered to be located at the inner lagrangian point L$_1$, at a distance $b_1$ from the center of the WD. Its position is computed using the Plavec
and Kratochvil approximation \citep{Plavec_Kratochvil1964},

\begin{equation}
    \frac{b_1}{a} = 0.500 -0.227\log_{10}\left(\frac{M_{\text{MS}}}{M\WD}\right) \, , 
\end{equation}

\noindent where $a$ is the separation between the two stars that can be computed from simple Keplerian dynamics as,

\begin{equation}
    a = \sqrt[3]{\frac{GMP^2}{4\pi^2}} \, ,
\end{equation}

\noindent with $M = M_{WD} + M_{MS}$ the mass of the binary system. 

The inner radii of our accretion disks were set at a distance $R_{\text{in}} = \frac{a}{20}$ instead of in immediate contact with the WD. We selected this position for practical reasons so that the geometrical setting of the different models is more uniform and the different ejecta velocities have reached the homologous phase\footnote{It should be noted that close to the WD the $\alpha$-disk models break down as magnetic fields, radiation pressure, and boundary layers become dominant \citep{frank1992}, and it remains poorly understood despite decades of theoretical work \citep{Piro2004,Hertfelder2013,Belyaev2013}.}. Then, the hydrodynamic profiles (height $H$, density $\rho_c$ and temperature $T_c$ at $z=0$) are given by the Shakura-Sunyaev solution:

\begin{equation}
    H = 1.7\times 10^8 \alpha^{-\frac{1}{10}}\dot{M}_{16}^{\frac{3}{20}}M_1^{-\frac{3}{8}}R_{10}^{\frac{9}{8}}f^{\frac{3}{5}}\, ,
\end{equation}

\begin{equation}
    \rho_c = 3.1\times 10^{-8} \alpha^{-\frac{7}{10}}\dot{M}_{16}^{\frac{11}{20}}M_1^{\frac{5}{8}}R_{10}^{-\frac{15}{8}}f^{\frac{11}{5}}\, ,  
\end{equation}

\begin{equation}
    T_c = 1.4\times 10^4\alpha^{-\frac{2}{10}}\dot{M}_{16}^{\frac{6}{20}}M_1^{\frac{2}{8}}R_{10}^{-\frac{3}{8}}f^{\frac{6}{5}}\, ,    
\end{equation}

\begin{equation}
   f = \sqrt[4]{1-\sqrt{\frac{R_{\textit{in}}}{R}}}\, ,    
\end{equation}

\noindent where $\dot{M}_{16}=\frac{\dot{M}}{10^{16}} $ \msun $ $ $ \text{s}^{-1}$, $M_1=\frac{M\WD}{M_\odot}$ and $R_{10}=\frac{R}{10^{10}}$ cm. It is important to mention that these expressions provide the physical magnitudes at the center of the disk (i.e. $z=0$). Although the disk is considered isothermal along its height with $T(r,z)=T_c(r)$, the thin disk solution provides an exponentially decreasing density profile in the range $[-H,H]$ such that:

\begin{equation}
    \rho(R,z)=\rho_c(R) e^{-0.5\frac{z^2}{H^2}} \, .
\end{equation}

\noindent Lastly, the orbiting velocities of the accretion disks have been chosen to be Keplerian, i.e. $v_\phi = \sqrt{GM\WD/R}$, and their chemical composition as solar with mean molecular weight $\mu = 0.614$. See Figure \ref{fig:profiles_disk} for the profiles of the different accretion disks adopted in the models.

\subsection{White dwarf and companion star}

The main objective of this work is to simulate the interaction between the ejected material from the nova/supernova outburst and the accretion disk. For that reason, the WD is added to the simulation just as a point-like source of gravity\footnote{Although the WD  is taken only as a source of gravity, a complete description of the star is used by the SHIVA code to model the outburst and obtain the radial profiles for the ejecta.} that keeps the disk bound to the system. In the case of models RN1, RN2 and SNIa the mass of the WDs is close to the Chandrasekhar limit, since these are the typical values of RNe and type Ia supernovae. The companion star and its gravitational effects are not included. This is a justified approximation since the disk spans the region dominated by white dwarf gravity (from the WD to the lagrangian point L1) and the gravity of the companion star can be neglected during the collision phase of the interaction and slightly after it. Also, the collision takes place at much shorter timescales than the orbital period, which makes any possible tidal or disk precession effects variations negligible. Other works have also adopted only an attractor+disk in an axisymmetric framework to gain resolution and make feasible simulations \citep{Romanova2006,Sadowski_etal_2014}. Nevertheless, later in section \ref{sec:disruption} we discuss the potential chemical pollution of the companion star after the arrival of ejecta and disk material.

\section{Numerical Approach} \label{sec:numericalapproach}
\subsection{Axisymmetric Smoothed Particle Hydrodynamics}

The collision between the ejecta and the disk is a boundless expansive event where the disk is highly deformed and hydrodynamical instabilities appear at multiple scales. For that reason a 
Lagrangian meshless method like Smoothed Particle Hydrodynamics (SPH) has been used to carry out the simulations. In this work, we have used the state-of-the-art SPH code AxiSPHYNX \citep{garcia-senz2023, vurgun2025}, which is the axisymmetric version of the full 3D code SPHYNX \citep{Cabezon_2017} and we refer the interested reader to these works for details of the implementation.

The choice for axisymmetry originates from the cylindrical geometry of the WD-ejecta-disk system centered on the primary star axis. These 2D simulations can effectively cope with important physical processes such as shock formation, momentum transfer, and angular momentum redistribution. Nevertheless, the main shortcoming of this approximation is that it only allows for an approximate treatment of three-dimensional effects linked to the development of hydrodynamic instabilities and turbulence. However, in many cases, these effects do not significantly alter the overall outcomes in the dynamics of the disk-ejecta interactions.

The dimensional reduction boosts the resolution with relatively low number $N$ of SPH particles due to the scaling relation $N_{3D}\simeq N_{2D}^{3/2}$. For example, an axisymmetric simulation with $10^6$ particles equals a full 3D run of $10^9$ particles, providing the same spatial resolution but with     a factor $\sim 10^3$ reduced computational cost.
 
The combination of low densities with moderate and high temperatures involved in the collision makes it feasible to use a simple equation of state, consisting of an ideal gas of ions plus a radiation term, where the pressure $P$ and internal energy per unit of mass $u$, are calculated as,

\begin{equation}
    P = \frac{R_{gas}}{\mu}\rho T + \frac{\alpha_\gamma}{3}T^4 \, ,
\end{equation} 

\begin{equation}
    u = \frac{3}{2}\frac{R_{gas}}{\mu}T + \alpha_\gamma\frac{T^4}{\rho} \, ,
\end{equation}

\noindent where $R_{gas} = 8.317\times 10^7$ $\text{erg} \, \text{K}^{-1} \, \text{mol}^{-1}$ is the ideal gas constant and $\alpha_\gamma = \frac{4\sigma}{c} = 7.564\times 10^{-15}$ $\text{erg} \, \text{cm}^{-3} \, \text{K}^{-4}$
is the radiation constant.  

\subsection{Distributing the SPH mass particles}

Achieving the desired density profiles is a delicate task due to the randomness involved in traditional initial SPH configurations. Typical algorithms rely on probability distributions proportional to the corresponding density profile, such that particles are placed randomly throughout the domain and the density of the set-up gets more accurate as $N \to
\infty$.

However, this process presents some inconveniences. Firstly, random placing can result in irregular distributions with regions having clumped particles and others with empty gaps, which results in important density fluctuations. This is commonly addressed by the so-called relaxation of the initial model, in which the initial model is left to evolve until particles reach a more stable configuration \citep[see e.g.,][]{garcia-senz2020}. However, this is time-consuming  because the convergence process is usually slow. For that reason, we have devised a novel particle method to set the SPH particles, which is based on the uneven triangulation of the particle sample that does not require any further relaxation. The details of that method are described in the Appendix.  

\section{Results} \label{sec:results}

Here we present the results of the simulations, focussing on the survivability of the disk, the evolution of the temperature in both the ejecta and the disk, the impact of emerging hydrodynamic instabilities and the diverting effect caused by the disk on the geometry of the ejecta. Additionally, we discuss the observational consequences for several of these issues.

\subsection{Disk Survivability}
The survivability of the disk has been studied using energy arguments. Material exceeding the escape velocity threshold is expected to unbind from the WD and leave the disk. Thus, we measured the percentage of disk mass that remains orbiting and its evolution throughout the simulation.

\begin{figure*}[h!]
    \centering
    \includegraphics[width=0.8\linewidth]{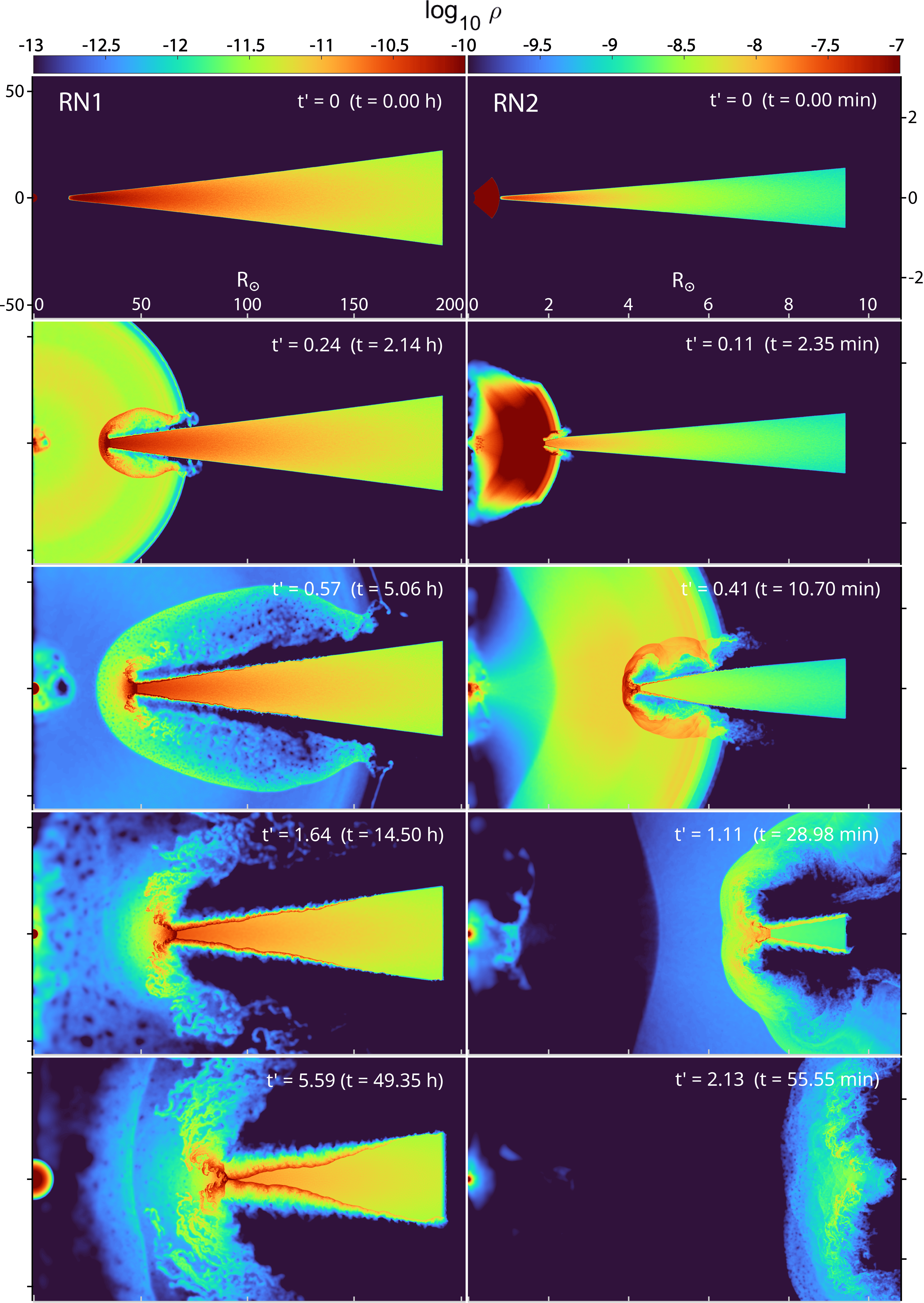}
    \caption{Snapshots of the density evolution (density in g\,cm$^{-3}$) at different times during the collision, displaying the distinct stages of the interaction with the disks, for models RN1 (left column) and RN2 (right). Axes display position (r,z) in solar radii, the same applies to all snapshots in figures \ref{fig:SnapshotsCN}, \ref{fig:maxtemprec}, \ref{fig:Turb} and \ref{fig:Screening}). The clumping of receding material around the WD position is not physical and is a result of gravity softening to avoid numerical instabilities for particles approaching $(r,z) = (0,0)$. Videos showing the full evolution are available as RN1.mp4, RN2.mp4 and RN2\_zoom.mp4.}
    \label{fig:SnapshotsRecurrent}
\end{figure*}

Figure \ref{fig:SnapshotsRecurrent} shows the density evolution of models RN1 and RN2, with snapshots taken at different times of collision. The difference in orbital periods ($P_{orb} = 454 $ and $P_{orb} = 5 $ days, respectively) has a huge impact on the outcome of the simulations. High values of $P_{orb}$ result in larger but less dense disks. It also causes the collision with the ejecta to occur farther from the WD, with the corresponding density reduction in the ejecta. However, according to the disk model adopted (Shakura-Sunyaev's thin disk) the total mass of the disk in model RN1 is far larger than that of model RN2 owing to its greater size. This explains the different results between RN1, where roughly 78\% of the disk survives, and RN2, where it is completely wiped away by the ejecta. These results can be understood by directly comparing the resulting mass and binding energy of the disks with the fraction of the ejecta impacting the disk:

\begin{equation}
 M_{disk}=\int_{\mathbf{V}}\rho \, dV=\int_0^{2\pi}\int_{R_{in}}^{R_{out}}\int_{-H(r)}^{H(r)} \rho(r,z)\, r \, dz \, dr \,d\phi \, ,  
\end{equation}

\begin{equation}
    \tilde{M}_{\text{ejec}} = M_{\text{ejec}}\,sin\, \frac{\theta_{\text{disk}}}{2} \, ,
\end{equation}

\noindent where $\tilde{M}_{\text{ejec}} $ is the effective ejected mass, an estimate of the fraction of the mass that actually impacts the disk, which has been considered to be all matter in the spherical sector $|\theta| < \frac{\theta_{\text{disk}}}{2} $. Analogously, we can do the same for their respective energies:

\begin{equation}
\begin{split}
 E_{disk}=E^{disk}_{kin}+E^{disk}_{pot}+E^{disk}_{int}\approx\\\approx \frac{1}{2}E^{disk}_{pot}=\frac{1}{2}\int_{\mathbf{V}}-G\frac{M\WD\rho}{r} \, dV  \, ,
 \end{split}
 \end{equation}

\begin{equation}
\tilde{E}_{\text{ejec}} \approx \tilde{E}^{ejec}_{kin} = E^{ejec}_{kin}\,sin\, \frac{\theta_{\text{disk}}}{2} = sin\, \frac{\theta_{\text{disk}}}{2}\int_{\mathbf{V}}\frac{1}{2}\rho v^2 \, dV  \, ,
\end{equation}

\noindent where $E_{kin}$, $E_{pot}$, $E_{int}$ are the kinetic, potential and internal energy contributions and $\tilde{E}_{\text{ejec}}$ is the effective energy of the ejecta (i.e. the energy carried away by $\tilde{M}_{\text{ejec}}$).

\begin{figure}[h!]
\centering    
\includegraphics[width=1\linewidth]{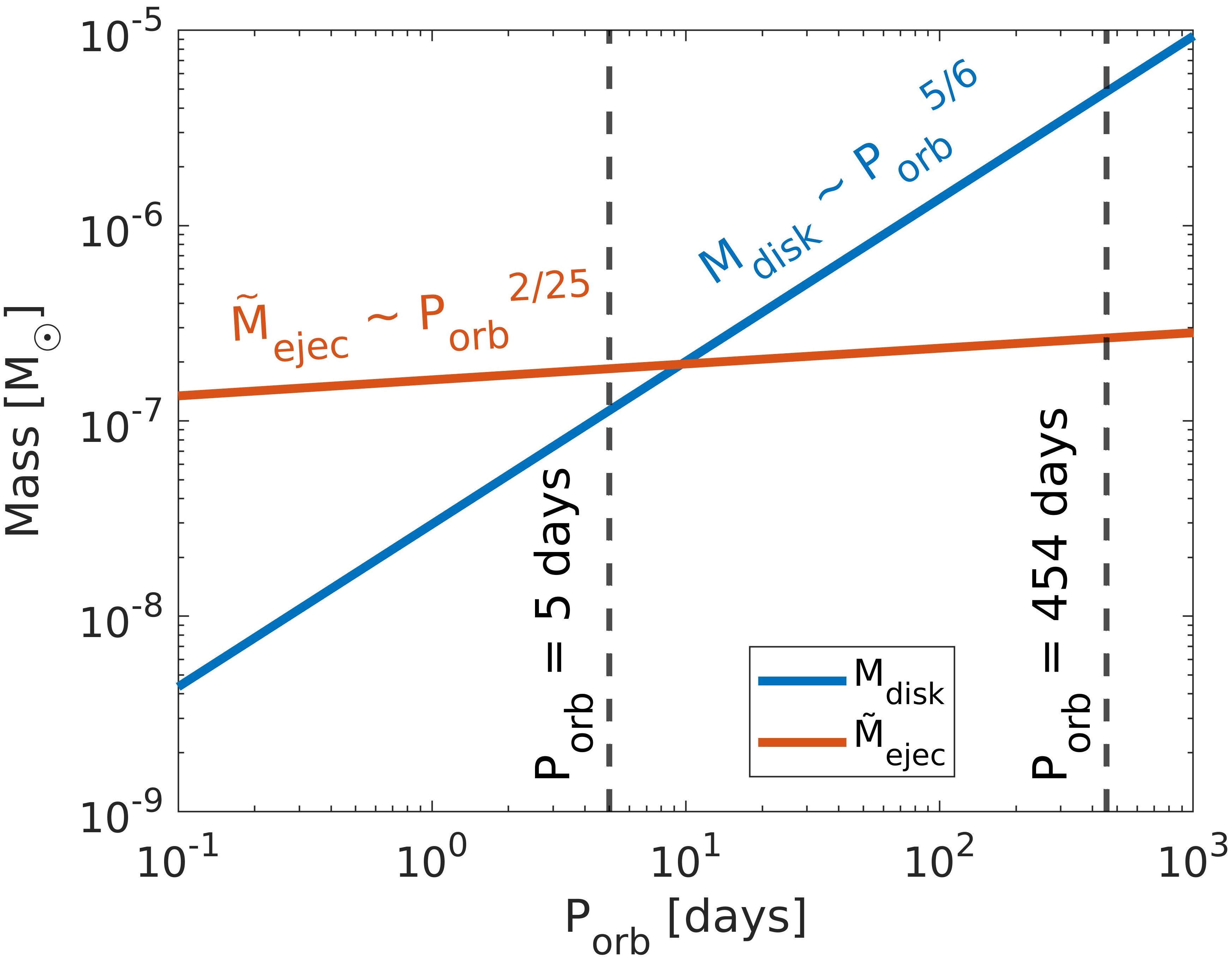}
\includegraphics[width=1\linewidth]{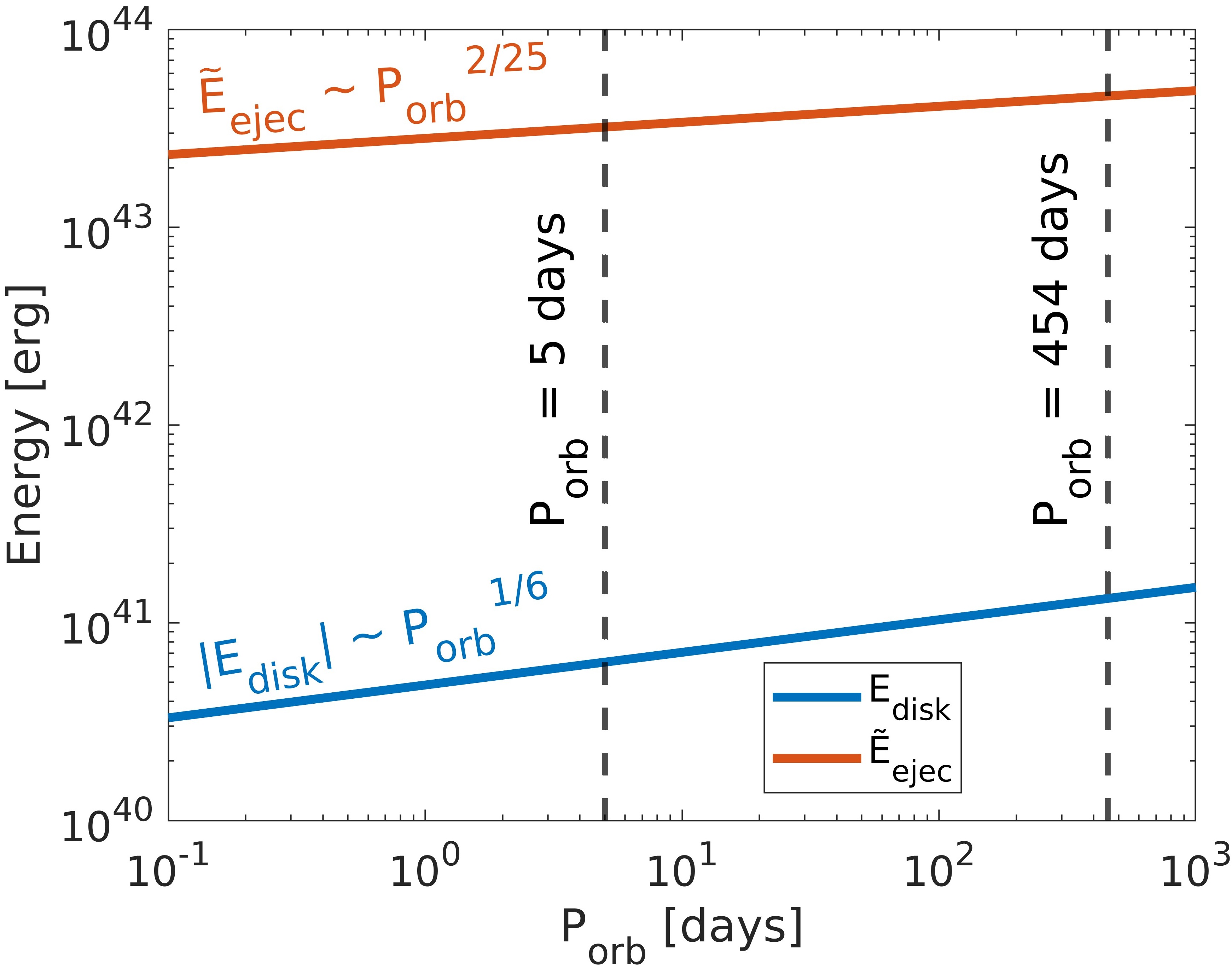}
    
    \caption{Masses and energy of the generated accretion disk and effective ejecta (for models RN1 and RN2) as a function of the orbital period $P_{\text{orb}}$. The disk has lower densities for larger values of $P_{\text{orb}}$ but its total mass increases. Nonetheless, the disk remains thin: the opening angle and, in consequence, $\tilde{M}_{\text{ejec}} $ barely increase. The second plot shows the absolute value of $E_{disk} $ since it mostly corresponds to negative potential binding energy.}
    \label{fig:massdependence}
\end{figure}

The first plot in Figure \ref{fig:massdependence} shows the dependence of these magnitudes on the orbital period $P_{\text{orb}}$. As shown, the disks formed in models RN1 and RN2 contain $4.84\times 10^{-6}$ and $1.13 \times 10^{-7}$ \msun $ $ respectively, while the effective ejected masses are $2.66\times 10^{-7}$ and $1.85\times 10^{-7}$ \msun, giving mass ratios of $18.2$ and $0.61$. Nevertheless, the result of the collision is dependent not only on the quantity of mass of both bodies but also on how strong the gravitational bound is. The second plot in Figure \ref{fig:massdependence} shows the balance of the energies between the ejecta (whose energy is mainly kinetic) and the disk. It is important to note that $|E\disk|$ grows at a slower rate than the mass of the disk because the gravitational potential energy decreases as $V \propto r^{-1}$. Despite the disk surviving in model RN1 or being destroyed as in RN2, the energy carried away by the ejecta is always orders of magnitude higher. Thus, the key factor that affects the difference in the outcome is the total mass that the ejecta has to sweep away throughout its expansion. Although we have not tested other disk morphologies, it is good to remark that the deflective geometry of the disk does not favor kinetic energy transfer, as it efficiently redirects the flow rather than absorbing it head-on. This explains its survivability against an ejecta that is orders of magnitude more energetic. 

\begin{figure*}[h!]
    \centering
    \includegraphics[width=1.0\linewidth]
    {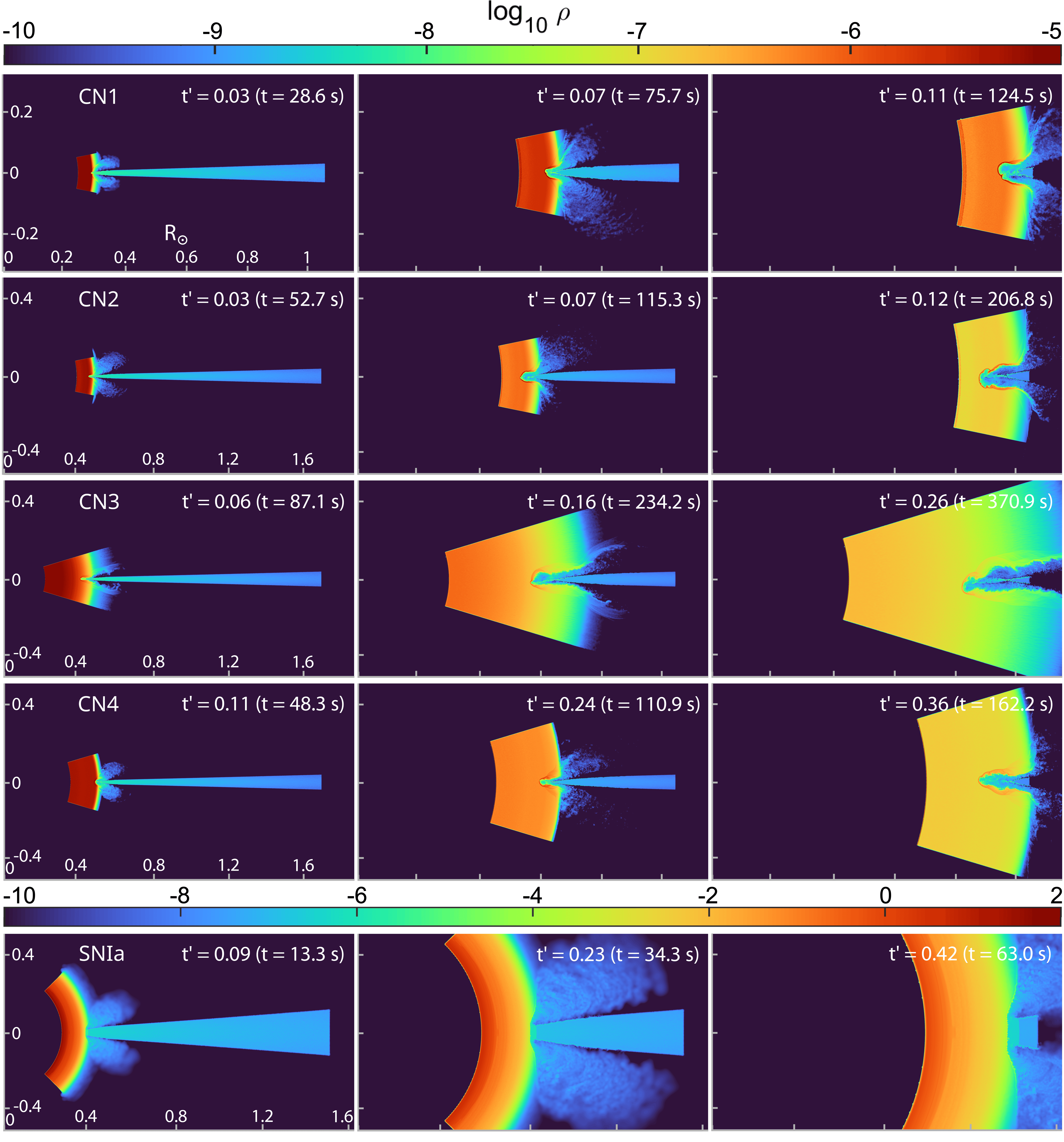}
    \centering
    \caption{Same as figure \ref{fig:SnapshotsRecurrent} for models CN1 to CN4 and the supernova model SNIa. Since only a reduced section of the ejectas are simulated, we have implemented reflective boundary conditions by adding ghost particles on both the sides and the innermost layer to prevent unphysical expansion beyond the domain cuts. Videos for these models are also available in the supplementary material.}
    \label{fig:SnapshotsCN}
\end{figure*}

Figure \ref{fig:SnapshotsCN} displays the density evolution for the rest of the CNe and the SNIa scenarios. Due to the enormous mass contrast between the disks and the ejecta in these models only a convenient outer region of the ejectas are simulated (see Table \ref{tab:Nparticles} in the appendix for details). As expected, the disks are easily wiped even by the outermost layers of the ejecta because they are significantly less massive than in the RNe models (due to the strong dependency with the orbital period $M_{disk}\sim P^{5/6}$ given by the Shakura-Sunyaev solution), while the mass of the ejecta is considerably larger. An extreme case is that of the SNIa explosion, where the whole WD is blown away.

By comparing models CN1 and CN2 we can test the effect of the WD mass on the simulations. Keeping the orbital period of the system constant while changing the value of $M\WD$ causes a different separation between the stars. The disk in model CN2 is formed in a wider system with a stronger gravitational pull from the more massive WD, making the disk larger and more dense than in model CN1. In particular, the disks extend to $1.05$ and $1.69$ \rsun,  containing $4.18\times10^{-11}$ and $1.03\times10^{-10}$ \msun $ $ respectively (see Table~\ref{tab:parameters1}). As seen in the snapshots, the latter case digs a larger hole which penetrates deeper into the ejecta due to the increased total mass in the second disk (Table \ref{tab:penetration} shows the depth of penetration of the disk in each model).

\begin{deluxetable}{c|ccccc}[h]
    \tablecaption{Penetration in the Ejecta\label{tab:penetration}}
\tablehead{ ... & CN1 & CN2 & CN3 & CN4 & SNIa }
\startdata
        Depth [\msun] & 1.12E-7 &  1.85E-7 & 1.81E-8 & 1.95E-7 & 1.99E-6\\  
\enddata
\tablecomments{The depth value is expressed as the mass of the shell contained between the deepest point affected by the disk and the outer edge of the ejecta.}
\end{deluxetable}

Models CN2 and CN3 in Table~\ref{tab:parameters2} have not too different expansion velocities of the ejecta but significantly differ in the amount of ejected mass. However, although the average velocities of the ejected material are similar, their value at the outer layers are slighlty different, achieving greater speeds in model CN2 due to the higher slope in its homologous profile (see figure \ref{fig:profiles_ejecta}). This, together with the larger dimensions of the disk in model CN3, results in a factor $\sim2$ difference in the time required for its destruction. In addition, the diagonal outflows present in the rest of the CNe models are absent because the low densities and velocities in this model produce a relatively soft collision which favors a more laminar dynamic in the interaction.

Lastly, the effect of the ejection velocities is tested by comparing models CN3 and CN4. The mean velocity is $\sim 3.15$ times higher in model CN4, and roughly a factor 10 in the kinetic energy of the ejecta. We can see that the increase in kinetic energy is directly reflected on the penetration in the ejecta, reaching a greater depth than model CN3 by approximately the same factor. It is important to remark that a more energetic ejecta does not necessarily mean a less relevant role played by disk: in this case the collision is more violent and the imprint left by the disk in the ejecta is greater precisely because of the higher velocities. Also, this scenario is energetically equivalent to model CN2, since the order of magnitude lost in mass is compensated by the increased kinetic energy, which is again reflected in the similar penetration depths.

Regarding the SNIa model, the higher accretion rate builds a disk with $1.42\times10^{-8}$ \msun, which is two orders of magnitude more massive than in the CNe models. Nonetheless, this is much smaller than the ejected material and does not offer any significant resistance to the supernova blast. Due to the high density ejecta, the disk can not penetrate it  significantly and is completely blown away. 

\begin{figure}[h!]
\centering
     \includegraphics[width=1.\linewidth]{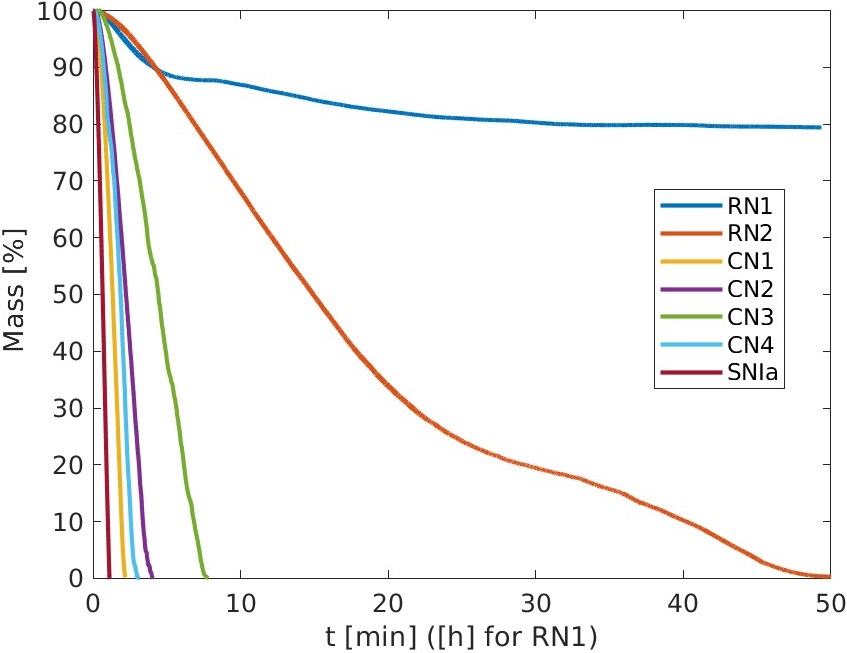}
     \includegraphics[width=1.\linewidth]{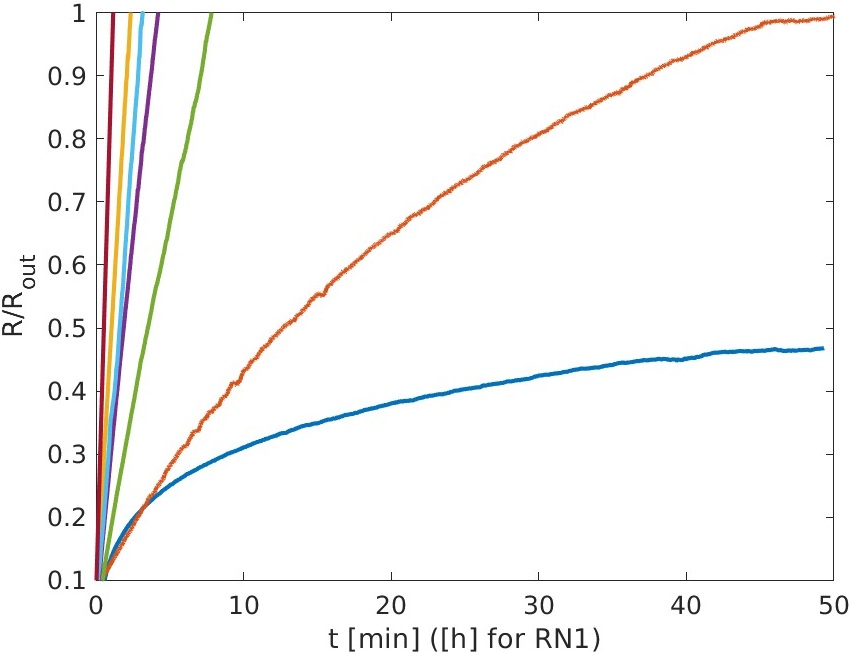}
    
    \caption{(Upper panel) Percentage of remaining mass bound in the accretion disk. (Lower panel) Position of the frontal shock of collision.}
    \label{fig:Evolution}
\end{figure}

As shown in Fig.~\ref{fig:Evolution}, and for all models, during the interaction the shock experiences a deceleration due to a combination of different effects:

\begin{enumerate}
     \item Increase of $H$: As the shock advances, the disk is wider and the contact surface grows.
     \item More mass: Although the disk becomes lighter as the distance from the WD increases, the mass swept continues to increase because the disk grows thicker and larger.
     \item Lower velocities: after some time, deeper layers of the ejecta take the lead in pushing against the disk, but at slower pace due to the initially homologous profile ( $v \sim r$).
     \item Ejecta expansion: Due to the expansion of the ejecta, its density and pressure keep decreasing, effectively reducing the net pushing.
\end{enumerate}

By studying the curves in Fig.~\ref{fig:Evolution} we can infer useful insight into the energetics of each scenario. In model RN1 the disk survives mostly unaffected and the shock stops moving forward halfway throught it. In contrast, in model RN2 the shock is able to reach the outer edge and fully wipe the disk. However, the flattening of the slope at the end of the curve indicates that this ejecta was close to the energetic threshold required for the destruction of the disk (i.e the shock would eventually stop advancing in a slightly longer disk, which would partially survive and keep its outer region unaltered).
The deceleration is negligible for the CNs and the SNIa scenarios since the mass carried by the ejecta is several orders of magnitude higher than in the disks; consequently, the disks are obliterated without much resistance. Only in model CN3 we barely appreciate the deceleration due to the combination of a slow ejection with low mass.

\subsubsection{Mirror symmetry across the Z$\pm$ half-planes}


  Mirror symmetry around the r-axis is not preserved over long periods in many of the calculated models, as shown in Figs. \ref{fig:SnapshotsRecurrent} and \ref{fig:SnapshotsCN}. Models where the disk offers little or no resistance to the ejected gas, such as the SNIa case, maintain the symmetry much better than those cases where the ratio of disk mass to impacting mass is larger, as in the RN1 model. The latter is reminiscent of a fluid crossing a cylindrical obstacle, which is believed to lose the mirror symmetry after crossing a critical point at Re $\gtrsim$ 50-100 \citep{herreros2020}, at this point the symmetric flow is unstable and any infinitesimal perturbation leads to a chaotic and turbulent asymmetric behaviour. Since typical Re numbers in our simulations are significantly higher, it is natural that these asymmetries develop, seeded by the numerical noise characteristic of SPH schemes\footnote{ The ring-like nature of particles in axisymmetric SPH codes makes axial geometry more prone to numerical noise than their 3D counterparts. This is because the mass of the SPH particles is that of a hoop and, therefore, the statistical fluctuation of a single particle in the SPH equations is higher. We have checked symmetry conservation by carrying out two Sedov tests in 2D and 3D. We obtained symmetry deviations in density $\sim 0.1\%$ and $\sim 0.01\%$, respectively.} (discretization, neighbour-particle search, etc) and also the non perfectly symmetric initial model (see Fig. \ref{fig:discretization}). Nonetheless, it should be noted that real accretion disks do not necessarily exhibit perfect symmetry of this type either.

\subsubsection{Numerical Convergence}

 In order to study the numerical convergence and explore the impact of resolution in the simulations, we have performed two additional variations of model RN1, with half and twice the number of SPH particles than the original calculation. The new models RN1-Low and RN1-High contain 0.5 and 2 million particles respectively and are distributed following the same profiles as in model RN1. All three variants show a very similar evolution and reach the same outcome, with some stochastic variance due to the unique particle distribution in each initial setup (see figures \ref{fig:RN1comparison} and \ref{fig:ComparisonRN1} in the appendix). Nonetheless, there is some difference between the models in the scale of the turbulent flows that are produced during the collision. Increasing the resolution reduces the interaction range of individual SPH particles, which affects the range of viscous forces. This allows reaching and further resolving the finer turbulent structure that develops in the flow before entering the viscosity-dominated scales \citep{cabezón2025}. 

\subsection{Maximum Temperature}

\begin{figure}[h!]
    \centering
    \includegraphics[width=1\linewidth]{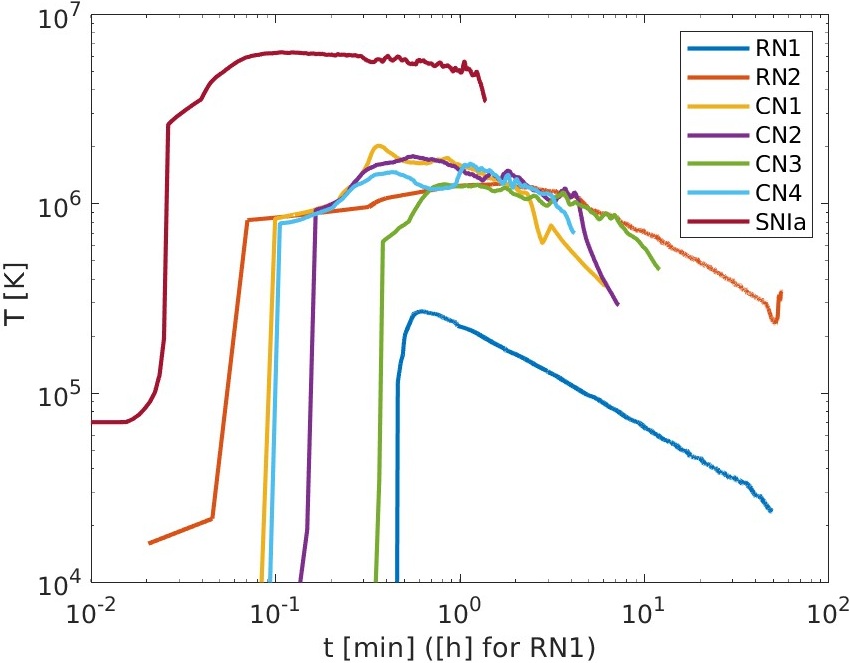}
    \caption{Time evolution of the maximum temperature reached by the shocked region in the disk.}
    \label{fig:maxtempevol}
\end{figure}

\begin{figure*}[t!]
    \centering
    \includegraphics[width=1\linewidth]{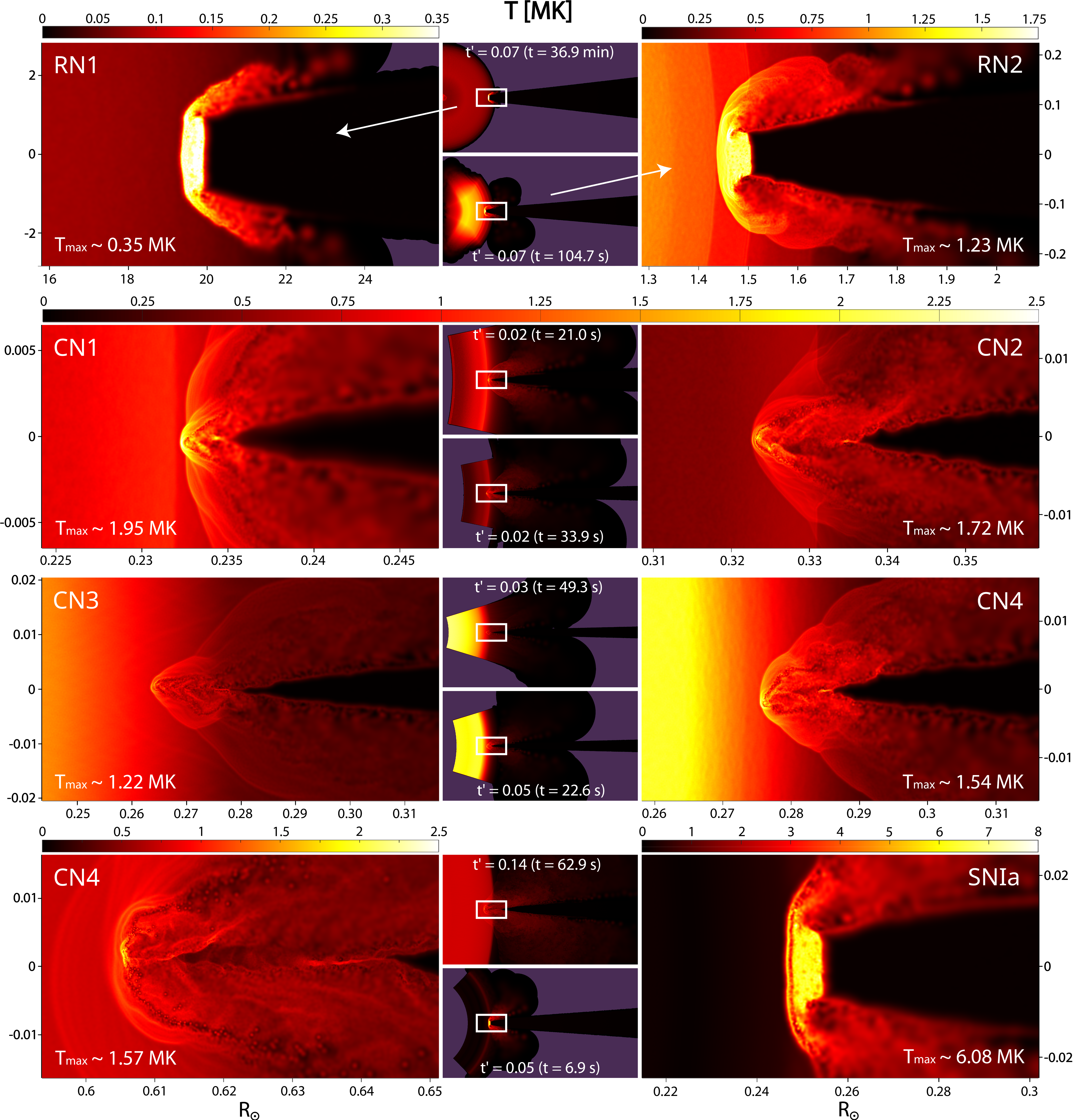}
    \caption{Close-up snapshots at the moment of maximum temperature of the collisions (horizontal axis corresponds to the position in solar radii and the colormap displays temperature in MK). Due to the high impact velocities, model CN4 reaches a second peak temperature later on due to the formation of hydrodynamical instabilities. Notice the different appearance of the shocks in models RN1, RN2 and SNIa when compared to those in the CNe models: the high accretion rates result in the formation of thicker and more robust disks that oppose increased resistance, forming this bubble shock morphology.}
    \label{fig:maxtemprec}
\end{figure*}

During the collision, viscosity is responsible for kinetic to internal energy transfer, resulting in the heating of the material. As seen in Fig. \ref{fig:maxtempevol} and Fig. \ref{fig:maxtemprec}  the maximum temperature is reached at the early stages of the collision, when the ejecta is not yet slowed and the impact takes place at maximum velocity against the inner and most dense region of the accretion disk.

Although the shock is somewhat violent and the material is strongly heated, new nuclear reactions are not able to develop during the collision because of the relatively low temperatures and densities within the event, $T < 10^7 \, \mathrm{K}$ and $\rho < 10^{-5} \, \mathrm{g}\,\mathrm{cm}^{-3}$ (nova), and $T < 10^8 \, \mathrm{K}$ and $\rho < 10^{5} \, \mathrm{g}\,\mathrm{cm}^{-3}$ (supernova). Nonetheless, the gas in the shocked material is heated to temperatures of the order of $\sim 10^6$ K, which is enough for a prompt X-ray signal due to those photons which escape freely, followed by an excess of UV radiation at longer times, once the radiation diffuses throughout the ejecta and thermalizes. This is reminiscent of the transient UV emission predicted to emerge during the collision of the SNIa ejecta with the companion star \citep{kasen2010}, which could have been observed \cite[e.g.][]{cao2015}. However, the observability of these specific UV and soft X-ray signals remains challenging for two main reasons. First, the whole collision takes place very fast in the first minutes of the outburst (hours in the RNe). Second, the signal can be hidden by the overwhelming luminosity of the whole event. Hence, whether the X-ray signal and the anomalous UV emission originated in the shock can be observed or not requires a comprehensive and detailed analysis beyond the scope of this work.

\subsection{Hydrodynamic Instabilities and Mixing}

\label{sec:hydroinstabilities}

The interaction between the nova ejecta and the accretion disk develops amid shocks, density discontinuities, and shear flows, which ultimately produce hydrodynamic instabilities and turbulence. Once formed, these instabilities accelerate the destruction of the disk and induce mixing between the material from the ejecta and the disk. The stirring effect of hydrodynamic instabilities generates turbulence, due to the huge Reynolds numbers involved ($Re\gtrsim 10^{14}$) in typical accretion disks \citep{frank1992}, even larger when considering the interaction with the higher velocity ejecta. Although hydrodynamic instabilities and, especially, turbulence are intrinsically three-dimensional phenomena, our axisymmetric approach can shed useful insights on the role played by these magnitudes.

\begin{figure*}[t]
    \centering
    \includegraphics[width=1\linewidth]{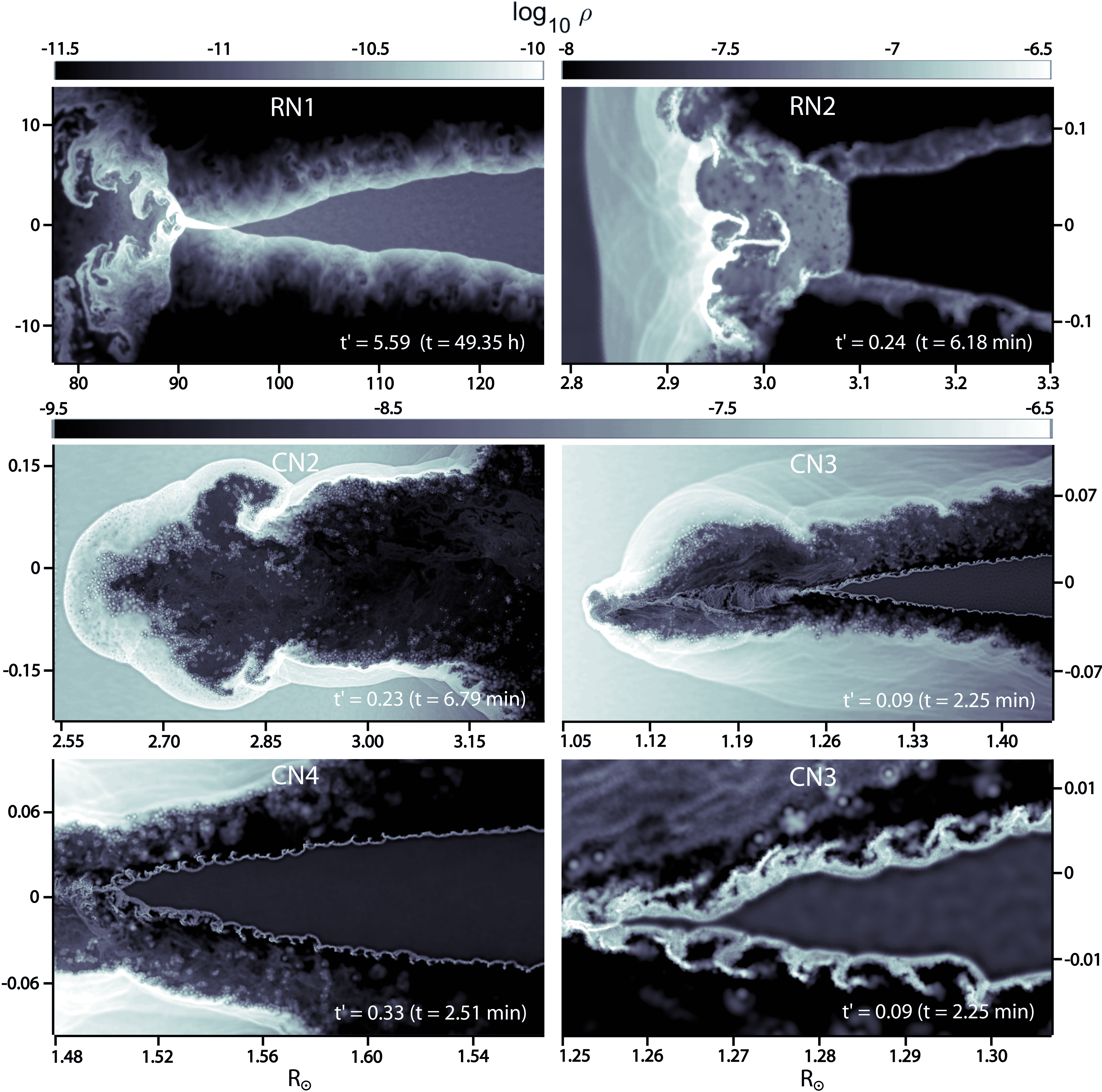}
    \caption{Density maps, in log scale, showing the formation of turbulent structures during the interaction.  }
    \label{fig:Turb}
\end{figure*}

The Kelvin-Helmholtz instability (KHI) is expected to take over during the first stages of the interaction because of the strong shear flow between the ejecta and the surface of the disk. The growth of KHI leads to vortex formation and corrugation of the disk surface. But soon, the high inertial forces that arise during the shock provoke the Richtmyer-Meshkov instability (RMI) to develop. The RMI appears whenever a shock hits the contact discontinuity that separates two fluids with different density \citep{richmyer1960,meshkov1969}. In astrophysics, the RMI typically appears during the interaction of supernova ejecta with circumstellar material \citep[e.g.][and references therein]{blondin2001}, which causes the growth of mushroom-like structures between the forward and reverse shock radii in supernova remnants (SNR). Both instabilities appear in many simulations presented in this work.

Figure \ref{fig:Turb} shows the density map around the interaction region between the ejecta and the disk of models RN1, RN2, CN2, CN3, and CN4 at different stages of the interaction. At these times, the disks still retain their identities, except that of model CN2 in which the disk is completely wiped and the image shows the gap produced in the ejecta. These plots show corrugated disk surfaces with clear signs of the KHI, plus superposed mushroom-like structures (see, for instance, the bottom surface of the disk in the snapshots of models RN2 and CN3). As the gas in the ejecta sweeps up more and more disk material, a reverse shock is formed, which travels down through the ejecta and heats the gas. The inertial force that arises from the deceleration of the shell facilitates the growth of convective instabilities. These are similar to the standard Rayleigh-Taylor structures, but referred to as the RMI in this context. A nice example can be seen in the second snapshot in Fig.~\ref{fig:Turb}, where the region between the reverse and forward shocks is filled with these mushroom-like objects. 

Figure \ref{fig:Turb} also shows a close-up of the ejecta-disk interface region for models CN3 and CN4. The surface of the disk is rippled with a number of small instabilities. The morphology of these instabilities is the result of combining the smooth vortexes of KHIs and the RMIs. In general, both KHIs and RMIs contribute to the fragmentation of the disk, accelerating its destruction. They also enhance the mixing of the material, which can have implications for the long-term evolution of the system. The incorporation of metal-enriched material into the sides of the disk can impact the evolution of the next nova cycle in the cases where the accretion disk survives, potentially changing the composition of the accreted material and consequently altering the next TNR. It is worth noting that hydrodynamical instabilities are responsible for the loss of mirrored symmetry in the horizontal axis: their presence in the shock can produce an asymmetric gap with an up or down preference that ends up in unbalanced diagonal outflows of material and ultimately into asymmetric nova remnants (see figure \ref{fig:Screening}). 

Although restricted to the axisymmetric plane, the collision with the disk produces a bow-shock in the ejecta, which is clearly visible in the different snapshots in Fig.~\ref{fig:maxtemprec}. Bow-shocks have been predicted to form during the early stages of the collision between the SNIa ejecta and the companion star in SD models of type Ia supernova explosions \citep{marietta2000, liu2013}. Our simulations show that the nova ejecta is also deflected by the accretion disk, giving rise to these geometries. But unlike the SNIa case, the bow-shock is more elongated and displays conical geometry only in the axisymmetric plane.

Interestingly, a reverse shock moving back through the ejecta is formed that heats the nova debris. According to Fig.~\ref{fig:maxtemprec} and depending on the particular nova scenario, the temperature may rise to $T_{max}\simeq 2$ MK. Although that temperature is below that observed in the reverse shock in current SNRs, $T\simeq 30$ MK \citep{decourchelle2000}, and lasts much less, it could transiently excite the spectral lines of some of the elements synthesized in the nova ejecta. Note that in the SNIa model no reverse shock is formed (last snapshot in Figure \ref{fig:maxtemprec}), although the temperature manages to briefly rise to $T_{\rm max}\simeq 6$ MK around the contact discontinuity in the front region of the disk.    

\subsection{Ejecta disruption, disk screening and chemical pollution of the companion star} 
\label{sec:disruption}

\begin{figure*}[h]
    \centering
    \includegraphics[width=1.0\linewidth]{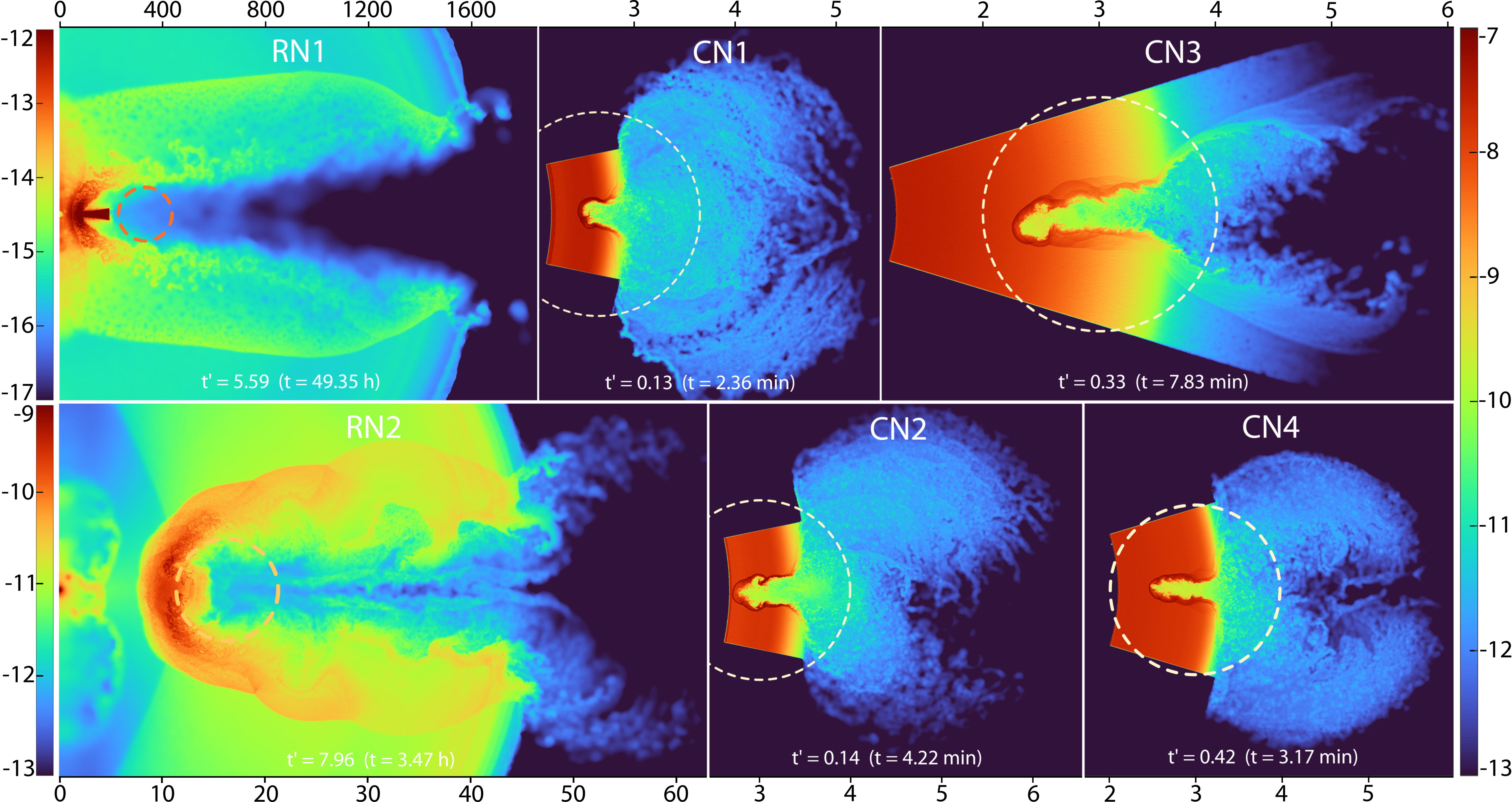}
    \caption{Density plot in logarithmic scale of the final remnants after the interaction has ceased. The dashed lines represent the location of the secondary star. As mentioned before, the companion in model RN1 is a red giant of $\sim110$ \rsun. Due to the orbital characteristics and accretion rate adopted in model RN2, the best fit for the companion is a subgiant between 4-6 \rsun. In our case we decided to use a 1 \msun $ $ and 5.23 \rsun $ $ star so that the system shared same geometric proportions than in model RN1. In the case of the CNe and SNIa models the companion is a Sun-like main sequence star with 1 \rsun $ $ and 1 \msun. }
    \label{fig:Screening}
\end{figure*}

One of the objectives of this work is to analyze the effect of the collision on the geometry of the ejecta. Figure \ref{fig:Screening} shows the expanded remnants when the interaction has virtually ceased. In the case of models RN1, RN2 and to some extent in CN3 the disk is able to break the ejecta, producing a two-lobed structure as the material is being deflected to the sides. In the rest of the scenarios, the disk digs a hole in the ejecta but the overall structure remains unaffected, especially in the SNIa scenario where the resistance opposed by the disk is practically negligible.

The hole created by the disk in the ejecta affects its geometry. Assuming that the nova shells are ejected spherically, the geometry is broken, first during the interaction with the disk and later with the companion star. In the case of type Ia supernova explosions it has been suggested that the hole produced by the companion star in the ejecta could be detected using spectro-polarimetric techniques \citep{kasen2004}. In a nova explosion the conical hole carved in the disk in the axial plane also extends all around the equator, in the azimuthal direction, which could also induce changes in the polarization state of the emitted light and be detected.

Apart from the geometry disruption and breaking of spherical symmetry, the deflection of the ejected gas has implications for the secondary star. As seen in Figure \ref{fig:Screening}, in the RNe scenarios the companion star lies inside the hole carved in the ejecta. Therefore, part of the ejecta that would have reached the star has been diverted to a new trajectory that does not collide with it. Hence, the accretion disk can act as a shield that screens the companion from the ejected nova material. However, in the case of the CNe and SNIa models, the disks are too light to produce any significant shielding. 

Previous calculations regarding the interaction of the nova ejecta with the companion star have found that a small fraction of the nuclear-processed material from the outburst can contaminate its atmosphere with CNO and intermediate-mass elements \citep{figueira2018}. This will alter the chemical composition of the surface layers, affecting the spectrum that would show peculiar traces of these nuclei. This has implications for the nucleosynthesis occurring during subsequent nova cycles, as the contaminated material will eventually be re-accreted onto the WD, forming a polluted and more metal-rich envelope. This altered composition will slightly alter the chain of nuclear reactions during the next TNR, potentially affecting the energetics, timescales, and nucleosynthetic yields of future outbursts. 

\begin{figure*}[t]
    \centering
    \includegraphics[width=1.\linewidth]{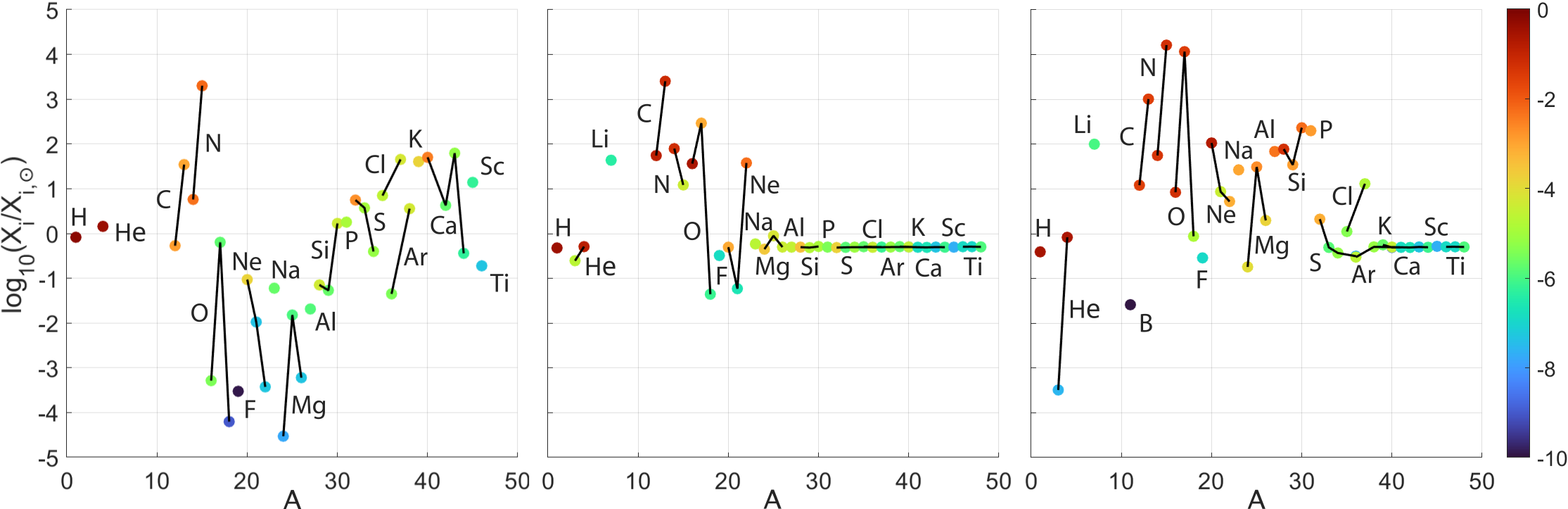}
    \caption{Overproduction factors in the ejecta relative to solar abundances, versus mass number for models RN1 and RN2 (left), CN1 and CN2 (center), CN3 and CN4 (right). The colormap shows the mass fractions of the different isotopes in log scale.}
    \label{fig:chemicalComposition}
\end{figure*}

 The nuclear reactions taking place during the early stages of the nova, i.e. during the TNR and early expansion, consist mostly in hydrogen burning predominantly via the CNO cycle, although proton capture reactions on heavier elements also occur. 
 Typical CNe ejecta from CO progenitors mainly produce isotopes from the CNO cycle, resulting in typical values of metallicity $Z \sim 0.1-0.5$ \citep{jose2016}. In the case of ONe WDs, apart from CNO, intermediate-mass elements in the Ne-Ca mass region are significantly produced. The resulting metallicities are slightly higher than those from CO WD progenitors (see Figure \ref{fig:chemicalComposition} for detailed chemical composition of the ejecta and \citealt{jose1998,jose2016} for an in-depth nucleosynthesis comparison between CO and ONe models). 

The conditions for the development of the TNR in RNe are characterized by shallow accreted layers with small quantities of dredged-up material, leading to an ejecta deficient in C and O when compared to CNe (the high mass of the WD along with high accretion rates shorten the duration of the accretion stage, and limit the mixing process operating at the base of the accreted envelope). Since the total mass ejected in the RNe outbursts is low and near-solar composition, the chemical pollution in the companion star will not alter significantly the overall metallicity of its outer layers \citep[see][for the recurrent nova U Sco]{figueira2025}. Such chemical pollution could be noticeable only for a handful of isotopes, highly overabundant in the nova ejecta (see Fig. \ref{fig:chemicalComposition}, left panel). Nonetheless, The screening of the disk will reduce significantly the amount of mass impinging on the secondary star, smoothing the overall pollution in the outer layers.

 We have made a rough estimate of the resulting chemical pollution for our novae models, with and without a disk, assuming that all material within a trajectory that intersects the secondary star will end up bound to it. As shown in Table \ref{tab:pollution}, the quantity of ejected mass reaching the star is significantly affected by the disk in both RN1 and RN2 models, with roughly $66\%$ and $55\%$ of the potentially polluting mass being deviated, reducing the pollution by a factor 3 and 2 respectively. 

As expected for the CNe models, the disk is unable to shield the companion star, causing the ejected material to pollute it indifferently of the presence of the disk. In these cases, the composition of the higher mass ejecta is enriched with both CNO and intermediate-mass elements, causing higher levels of pollution than in RNe, reaching abundances orders of magnitude higher than solar for certain nuclei such as \iso{C}{13} and \iso{N}{15}. However, the extent to which the material incorporated is transported and mixed in the companion star (either by diffusive or convective mechanisms), and to what depth is difficult to ascertain and out of the scope of this work, requiring a stellar model to analize the evolution of the secondary star. 

\begin{deluxetable*}{c|ccccccccccc}[t]
\tablecaption{Chemical pollution in the secondary\label{tab:pollution}}

\tablehead{
\colhead{Name} & \colhead{$Z$}  &\colhead{$M^{\text{ejec}}_{\text{trajectory}}$} & 
\colhead{$M^{\text{ejec}}_{\text{pollution}}$} &
\colhead{$\zeta$} &
\colhead{$M_{\text{pollution}}^{\text{disk}}$ } &
\colhead{$M_\text{pollution}$} &
\colhead{$Z_\text{pollution}$} & 
 \colhead{$X^{\text{\sout{disk}}}_{\text{\iso{C}{13}}}$} &
 \colhead{$X^{\text{disk}}_{\text{\iso{C}{13}}}$}  &
 \colhead{$X^{\text{\sout{disk}}}_{\text{\iso{N}{15}}}$} &  \colhead{$X^{\text{disk}}_{\text{\iso{N}{15}}}$} 
}

\startdata
        RN1 & 0.0186  & 2.639E-8 & 8.967E-9 & 0.660 & 3.393E-9 & 1.236E-8 & 0.0171 & 3.081E-5 & 2.916E-5 & 1.983E-5 & 8.847E-6\\
        RN2 & 0.0186  & 2.639E-8 & 1.200E-8 & 0.545 & 6.937E-9 & 1.894E-8 & 0.0167 & 3.081E-5 & 2.945E-5 & 1.983E-5 & 1.076E-5\\
        CN1 & 0.1066  & 7.405E-6 & 7.405E-6 & 0 & 5.106E-12 & 7.405E-6 & 0.1066 & 3.054E-2 & 3.054E-2 & 1.835E-5 & 1.835E-5\\
        CN2 & 0.1066  & 5.482E-6 & 5.482E-6 & 0 & 8.811E-12 & 5.482E-6 & 0.1066 & 2.543E-2 &  2.543E-2 & 1.580E-5 & 1.580E-5\\
        CN3 & 0.1147  & 5.429E-7 & 5.429E-7 & 0 & 9.269E-12 & 5.429E-7 & 0.1147 & 1.507E-3 & 1.507E-3 & 2.672E-3 & 2.672E-3\\
        CN4 & 0.1147  & 5.429E-7 & 5.429E-7 & 0 & 8.880E-12 & 5.429E-7 & 0.1147  & 1.507E-3 & 1.507E-3 & 2.672E-3 & 2.672E-3\\  
        \hline
        \textit{Solar} & 0.0139 & ...&... &... &... &... & ...& 2.831E-5 & 2.831E-5 & 3.180E-6& 3.180E-6\\
\enddata
\tablecomments{$Z$ is the mean metallicity of the ejecta. $M^{\text{ejec}}_{\text{trajectory}}$ is the ejected mass initially en route to the companion star, and $M^{\text{ejec}}_{\text{pollution}}$ is the actual ejected mass reaching the star after the collision with the disk and contributing to chemical pollution. $\zeta$ is the screening factor calculated as ($M^{\text{ejec}}_{\text{trajectory}}$-$M^{\text{ejec}}_{\text{pollution}}$)/$M^{\text{ejec}}_{\text{trajectory}}$. $M^{\text{disk}}_{\text{pollution}}$ is the mass with origin at the disk that collides with the companion and thus $M_{\text{pollution}}$ is the total mass that reaches the star with average metallicity $Z_\text{pollution}$. $X^{\text{\sout{disk}}}_{\text{\iso{C}{13}}}$ and $X^{\text{disk}}_{\text{\iso{C}{13}}}$ are the resulting abundances of \iso{C}{13} in the outer $10^{-5}$ \msun $ $ layer of the companion star, assuming that the material has been completely diffused and mixed evenly in it, without and with the disk, respectively. The last two columns are analogous, but for \iso{N}{15}.}
\end{deluxetable*}

\section{Conclusions} 
\label{sec:conclusions}

The interaction between the nova/supernova ejecta and the accretion disk has been studied by carrying out seven different simulations presented in this work, exploring the impact of varying parameters such as the orbital period, the WD mass, the ejected mass or the velocity profile of the ejecta. The axial symmetry of the adopted initial scenario allows to simulate these events with an unprecedented resolution, providing a clear picture of the physics of the interaction and its consequences. The study complements those by \cite{figueira2018,figueira2025},  performed in three dimensions but at much lower resolution, and confirms some of their findings related to the survivability of disks or the possible contamination of the external layers of the companion star with nova material. In the following paragraphs we summarize the main results and conclusions:

\begin{itemize}

    \item{The fundamental magnitude affecting the survivability of the disks is the mass contrast between the ejecta and the disk. When this mass-ratio is of the order of one or larger the disk will resist the collision to some extent. Among our simulations, only in our first model RN1 (resembling the RS Oph system) the disk survives the collision and is not completely destroyed, with roughly 78\% integrity after the interaction. This is the only model where the disk is more massive than the ejecta, which is a result of a combination of the low mass ejected characteristic of RNe and a large wide massive disk emerging from the high accretion rate and orbital period (see Tables \ref{tab:parameters1} and \ref{tab:parameters2}).}
    
    \item{In the scenarios where the disk mass is comparable with the ejecta (in our case, only models RN1 and RN2) we have observed clear disruption and deformation of the initial spherical geometry of the ejecta. When the disk is significantly resisting the impact, a bow shock forms in the ejecta that diverts and splits it into a two-lobed shape with a ring-like hole in the horizontal axis (see figure \ref{fig:Screening}). In contrast, models CN1 to CN4 and the SNIa model have shown only low-depth gaps in the outer layers of the ejecta. The ejecta splitting in models RN1 and RN2 produces a big gap which has important implications regarding the interaction with the companion star: the disk acts as a shield to the secondary and a significant fraction of material is deviated, decreasing the total mass impacting the star. Our results show screening factors of $66\% $ and $55\%$ respectively, which implies reduced chemical pollution of the companion star, with abundances of certain key isotopes (e.g. \iso{C}{13} and \iso{N}{15}) reduced by factors $\sim 1.5 - 2$.}

    \item{During the disk sweep, a shock wave is born which detaches from the contact discontinuity and travels  through  the disk. The immediate collision and the forward shock heat the material, producing high temperatures,  of the order of $10^6$ K,  which may result in prompt X-ray emission followed by an excess of UV emission once the photons diffuse throughout the ejecta and thermalize. Similarly, a reverse shock forms which moves across the ejecta (see the second snapshot in Fig. \ref{fig:Turb}), raising the temperature above $10^6$ K, potentially producing emission lines from various excited nuclei present in the ejected plasma. However, this needs to be confirmed by more detailed simulations that include radiative  transport. Between the forward and the reverse shock and around the contact discontinuity there is room for the development of hydrodynamic instabilities, especially the RMI, as shown in Fig.~\ref{fig:Turb}. Nonetheless, the observation of these X-Ray and UV signatures remain uncertain due to the high luminosity of the whole event and the short duration of the emission. 
    
    The hole carved in the shocked ejecta could also have observational consequences that merit exploration in future work. For example, the axial gap left in the ejecta in models RN1 and RN2 breaks the spherical symmetry and induces altered polarization in the emitted light which could be observed, as similarly suggested for the hole left by the presence of the companion star in SD models of type Ia supernova explosions \citep{kasen2004}. However, this raises the question of whether the gap is going to endure the expansion and the interaction between the remnant and the circumstellar medium. We plan to follow the long-term evolution of these systems in a future paper, and study the possibility of the gap to be closed or refilled with surrounding material, affecting the geometry and observable characteristics of the remnants after hundreds of years of expansion.}
\end{itemize}

\begin{acknowledgments}
This work has been partially supported by the Spanish MINECO grant PID2023-148661NB-I00, by the E.U. FEDER funds and by the AGAUR/Generalitat de Catalunya grant SGR-386/2021. This article benefited from discussions within ChETEC-INFRA (EU project no. 101008324).
\end{acknowledgments}

\bibliographystyle{aasjournal}
\bibliography{sample7}

\appendix

\section{Initial Setup. Triangulation and Relaxation}

The method used in this work aims to provide high-quality initial models, with more uniform density distributions and avoiding the relaxation process. We have used a 2-dimensional delaunay triangulation algorithm \citep{Engwirda-Darren}, this approach constrains inter-particle distances depending on density:

\begin{equation}
   l(r,z) \propto \left(\frac{m}{\rho(r,z)}\right)^\frac{1}{2} 
\end{equation}

Then, a 2D surface triangulation is performed with l as the target length of the edges of the triangles. After triangulation has been performed, the particles are placed at the vertex of the generated mesh.

\begin{figure}[h!]
\centering
\includegraphics[width=0.5\linewidth]{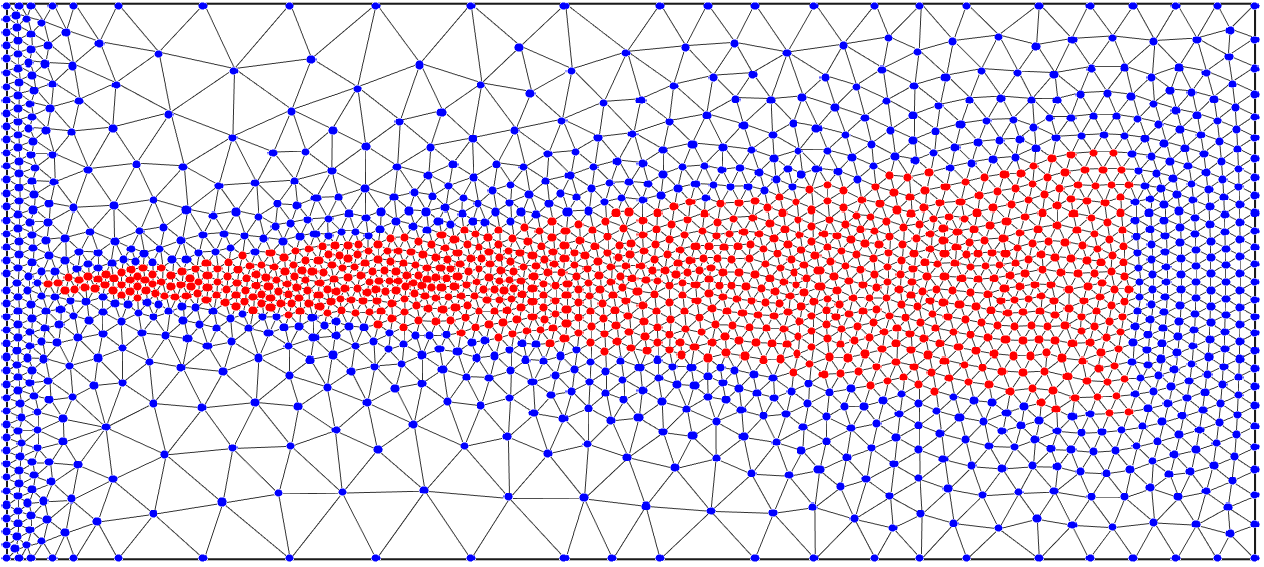}
    \caption{Low-resolution triangulation example of an accretion disk. Only the red particles laying inside the disk domain [$R_{in},R_{out},-H(R),H(R)$] are kept.}
    \label{fig:discretization}
\end{figure}

This method allows for having particles with constant mass, or with a certain mass profile. In any case, it is important to recall that in an axisymmetric framework SPH particles represent 3D rings centred around the axis, and for constant density the mass of these rings grows proportional to its perimeter, i.e. $m \propto r$. Thus, a more correct expression of the target length $l$, including an arbitrary profile for the masses of the particles:

\begin{equation}
 l(r,z) \propto \left(\frac{m(r,z)}{r\cdot\rho(r,z)}\right)^\frac{1}{2} 
\end{equation}

\begin{figure*}[t]
\centering
\includegraphics[width=0.44\linewidth]{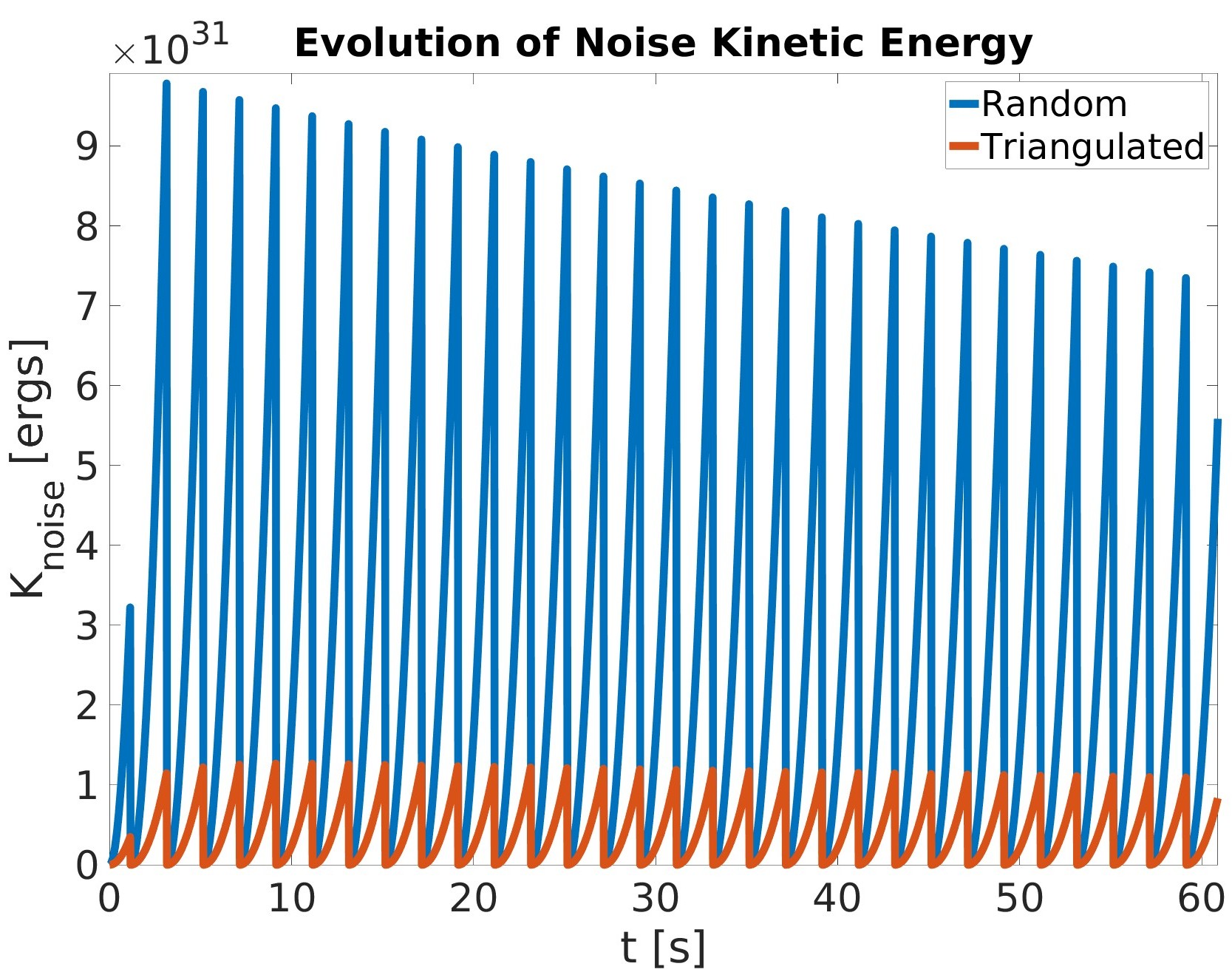}
\includegraphics[width=0.44\linewidth]{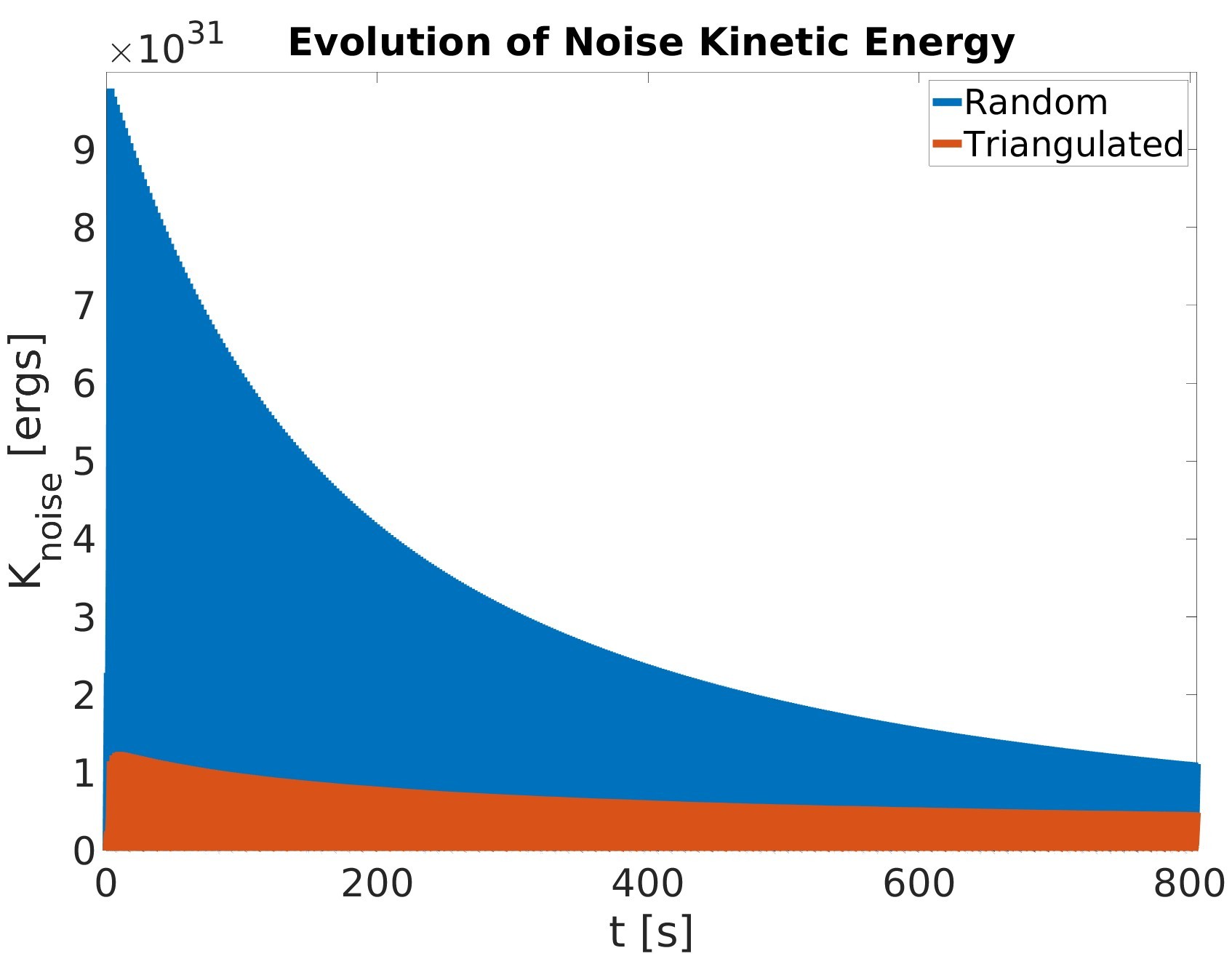}
\includegraphics[width=0.44\linewidth]{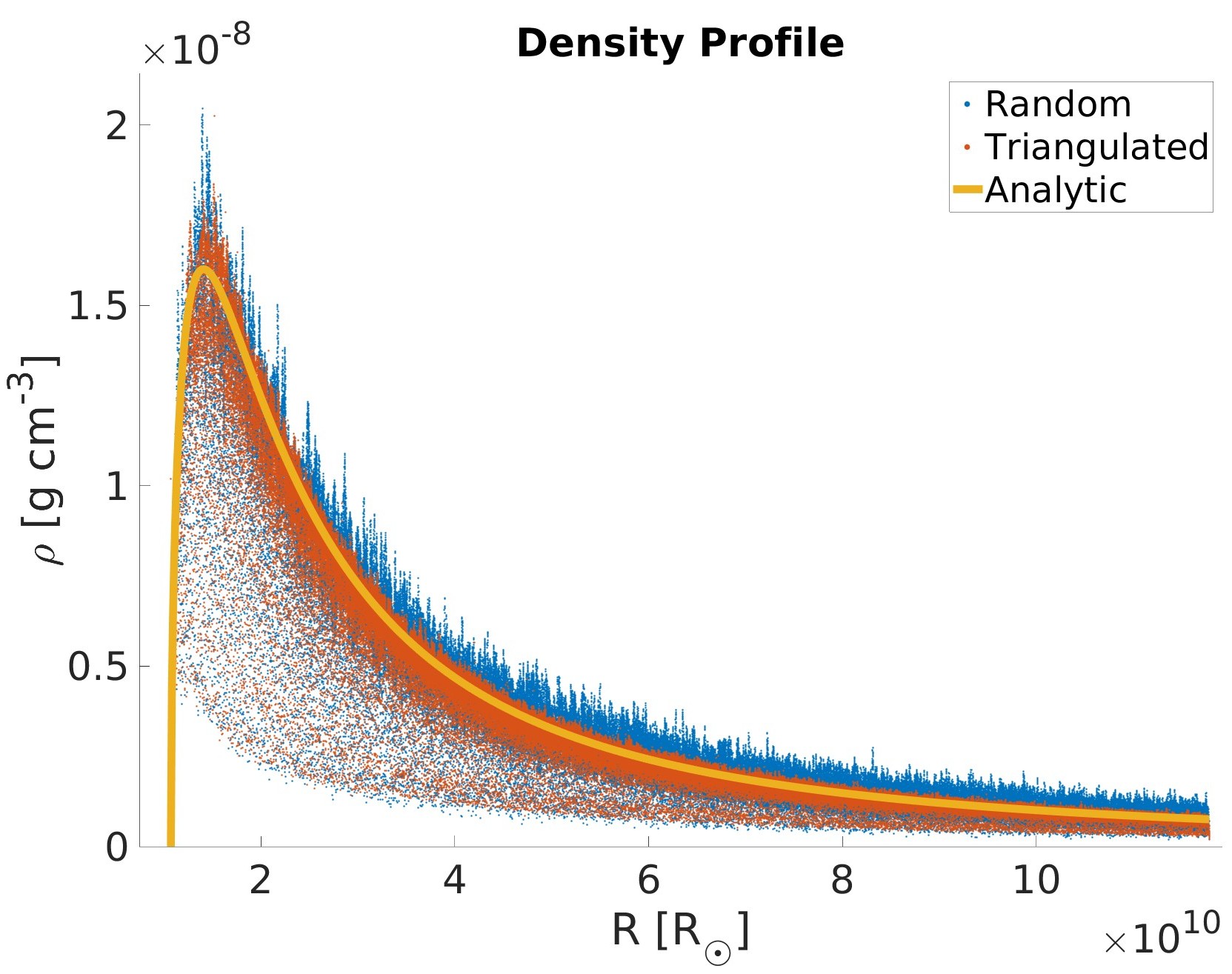}

    \caption{(Upper left pannel) Relaxation process, which consists in resetting the velocities every 100 iterations with the aim of reducing kinetic energy noise. (Upper right pannel) Complete relaxation process for $t$ several times the collision time. (Lower pannel) Post-relaxed density profile of the accretion disk in models CN2, CN3 and CN4.}
    \label{fig:relax}
\end{figure*}

  This method has many advantages compared to a random distribution. The triangulation ensures that particles do not overlap and are placed at similar distances from all their neighbors, and, in addition, this ordering provides straightforwardly a stable initial configuration avoiding the several iteration-steps needed for achieving a post-relaxed random distribution(see figure \ref{fig:relax}). In consequence, the resulting initial density profiles are less noisy and more similar to the original analytic profiles.

The proportion of particles distributed between the disk and the ejecta is mainly determined by the ratio of their masses. However, in some scenarios the ejected material is much more massive, which makes it impossible to have good resolution in the disk without enormously increasing the total number of particles. For that reason, only in the RNe models (where $M_{ejec}$ is relatively low) we have kept constant the mass of the SPH particles (see Table \ref{tab:Nparticles}). In the rest of the cases, the ejecta has been built with varying masses:

\begin{equation}
    m\ejec(r) \propto \rho(r)^\eta
\end{equation}

\noindent with $0\leq\eta\leq1$ as a parameter to control the extent to which the particles follow the original density profile. Also, we have opted to save computational resources by including only the material that is able to interact with the accretion disk during the collision; this means that only a conic fraction of the initially spherical ejecta and up to a reasonable depth is actually simulated.

\begin{deluxetable}{c|cccccc}[h!]
\tablecaption{Simulation data\label{tab:Nparticles}}
\tablehead{
\colhead{Name}  & \colhead{ $N_{ejec}$ } &
\colhead{$N_{disk}$} & \colhead{$\eta$} & \colhead{Depth [\msun]} &
\colhead{\% mass} &\colhead{Angle}
}
\startdata
         RN1  & 0.18 & 0.82 & 0 & 1.18E-6 & 100 & 180 \\
         RN2  & 1.20 & 0.43 & 0 & 1.18E-6& 100 & 80\\
        CN1 & 0.81 & 0.18 & 0.75 & 1.97E-7& 0.08 &24 \\
        CN2 & 0.81 & 0.48 & 0.75 & 1.97E-7& 0.08 & 24\\
        CN3  & 0.83 & 0.48 & 0.7 & 2.93E-7& 1.3&34\\
        CN4 & 0.83 & 0.48 & 0.7 & 2.93E-7 & 1.3&34 \\
         SNIa  & 1.44 & 0.18 &  0.7 & 1.00E-1 & 7.35&90 \\
\enddata
\tablecomments{$N_{ejec}$ and $N_{disk}$ are the number (in millions) of SPH particles in the ejecta and the disk. "Depth" and "\% mass" indicate up to which ejecta layers are included in the simulation, for example, only the outer 0.1 \msun $ $ are included in the SNIa scenario which supposes 7.35\% in mass of the total ejected material. The last column is the maximum angle simulated in the ejecta. }
\end{deluxetable}

\section{Convergence tests}

 We have carried out a convergence test by performing two additional simulations of model RN1, with half and twice the number of SPH particles. We opted for a factor 2 decrease/increase in resolution, such that the lower and higher resolution models differ a factor 4, corresponding to an effective factor of 8 when accounting for the 2D axisymmetric to 3D relation, which is already a substantial difference to observe the impact of resolution in the simulations. A test with a much larger number of particles would be challenging because of the large computational time required.

\begin{figure}[h!]
     \centering
     \begin{subfigure}[b]{0.48\textwidth}
         \centering
         \includegraphics[width=\textwidth]{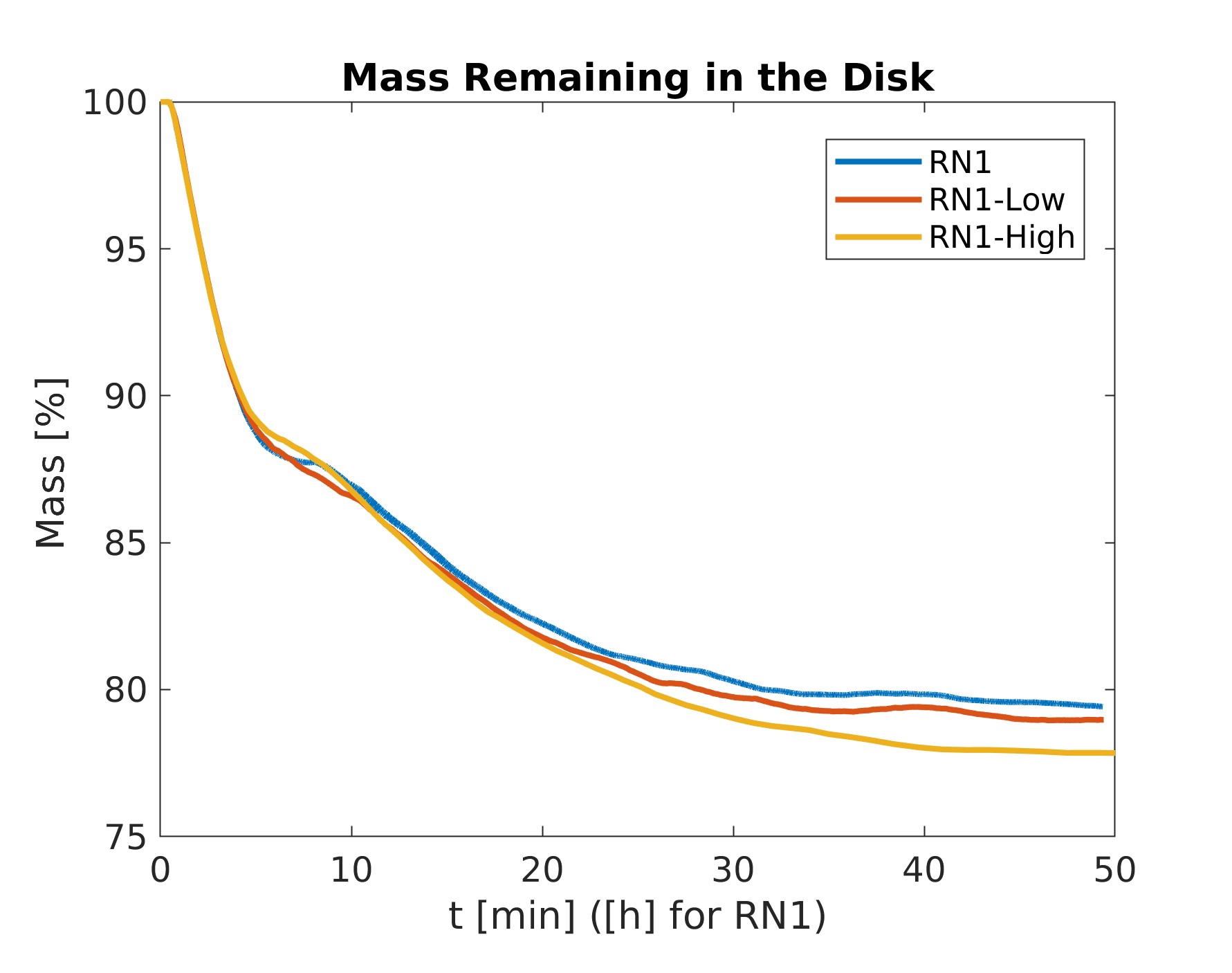}
         \label{fig:RN1Mass}
     \end{subfigure}
     \hfill
     \begin{subfigure}[b]{0.48\textwidth}
         \centering
         \includegraphics[width=\textwidth]{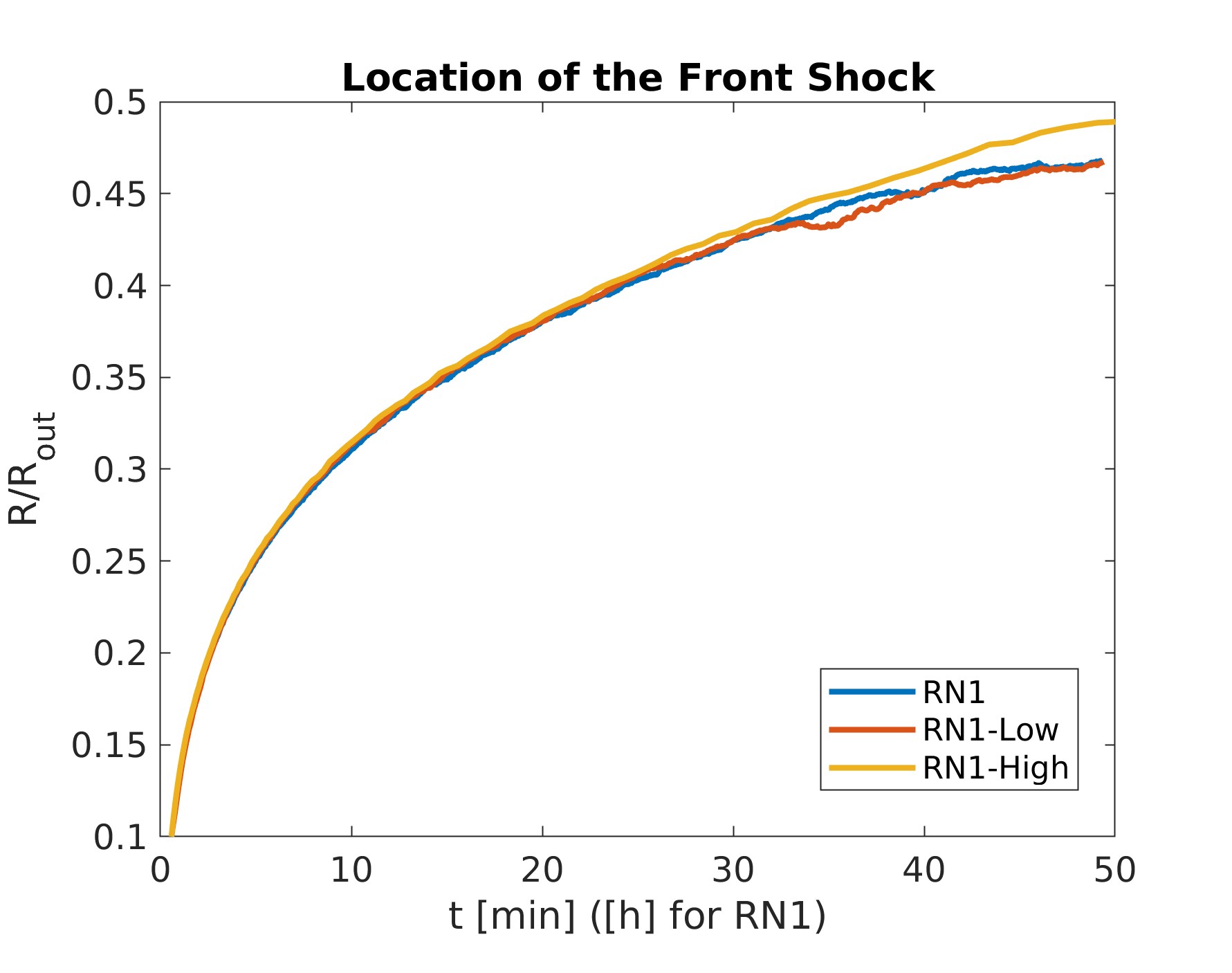}
         \label{fig:RN1Shock}
     \end{subfigure}
     \hfill
     \caption{Same as figure \ref{fig:Evolution} for the three variations on resolution of model RN1}
     \label{fig:ComparisonRN1}

\end{figure}

\begin{figure}
     \centering
     \begin{subfigure}[b]{0.6\textwidth}
         \centering
         \includegraphics[width=\textwidth]{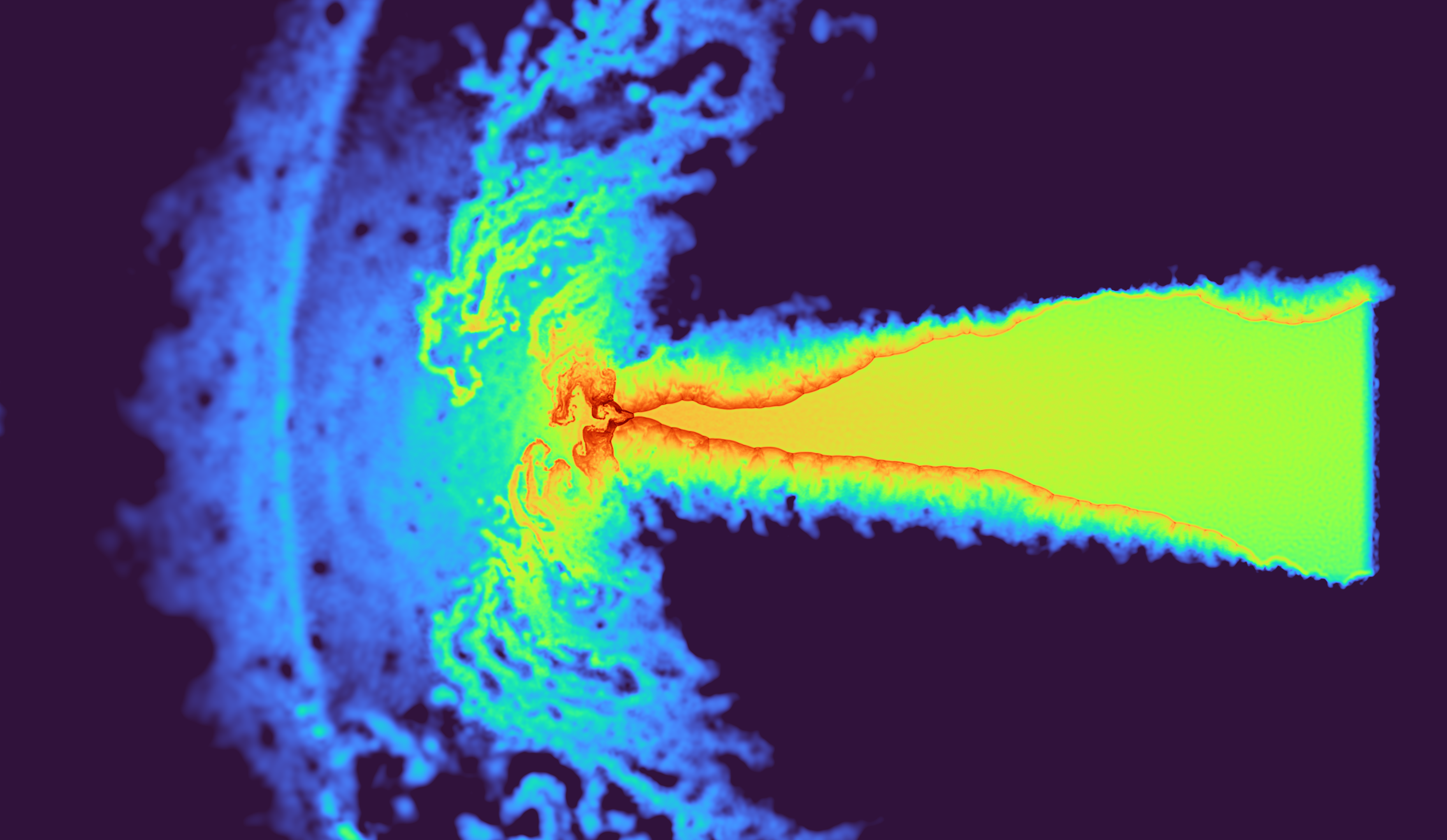}
         \caption{RN1-Low: N = 500,000 particles}
         \label{fig:RN1Low}
     \end{subfigure}
     \hfill
     \begin{subfigure}[b]{0.6\textwidth}
         \centering
         \includegraphics[width=\textwidth]{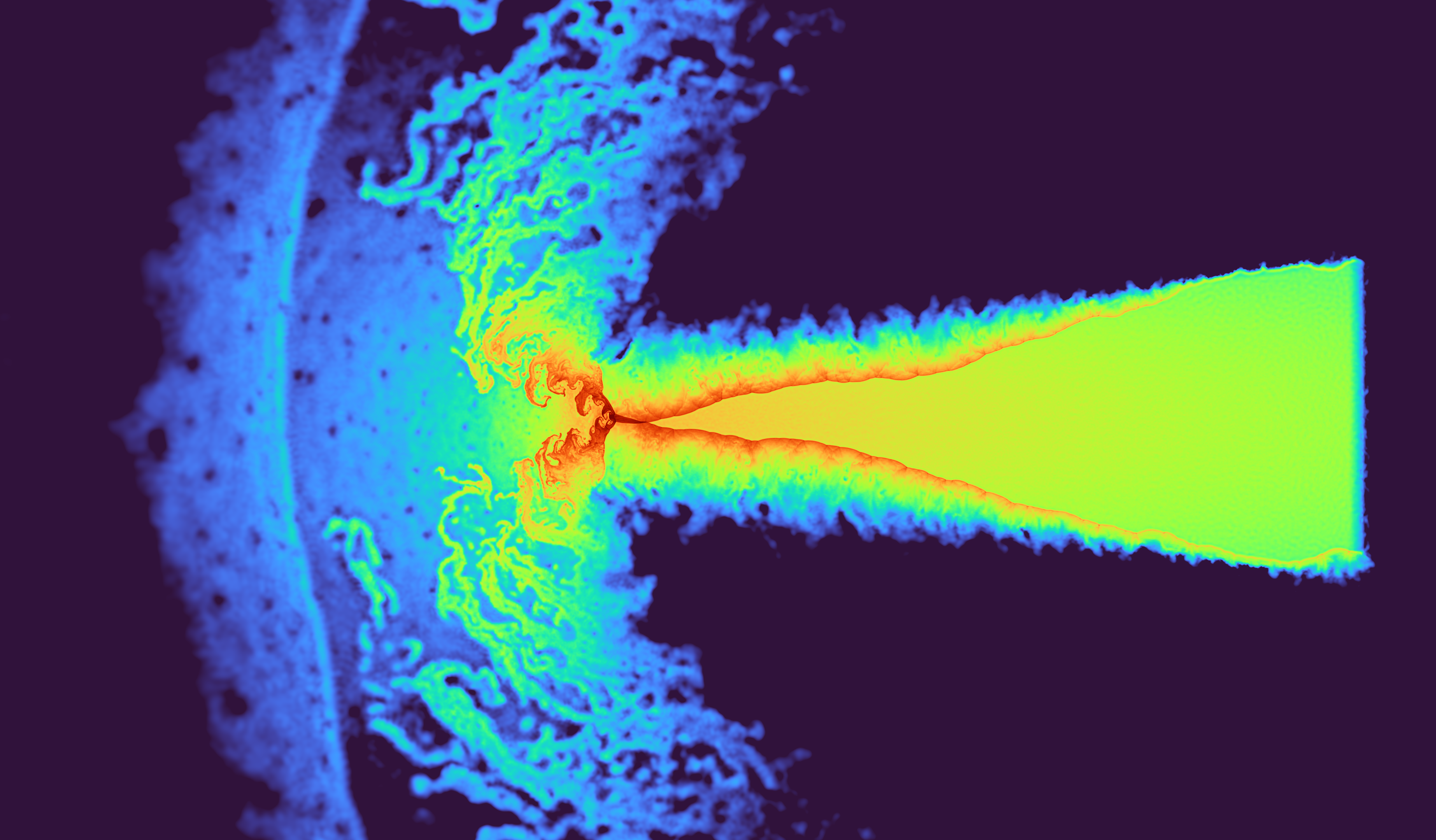}
         \caption{RN1: N = 1,000,000 particles}
         \label{fig:RN1}
     \end{subfigure}
     \hfill
     \begin{subfigure}[b]{0.6\textwidth}
         \centering
         \includegraphics[width=\textwidth]{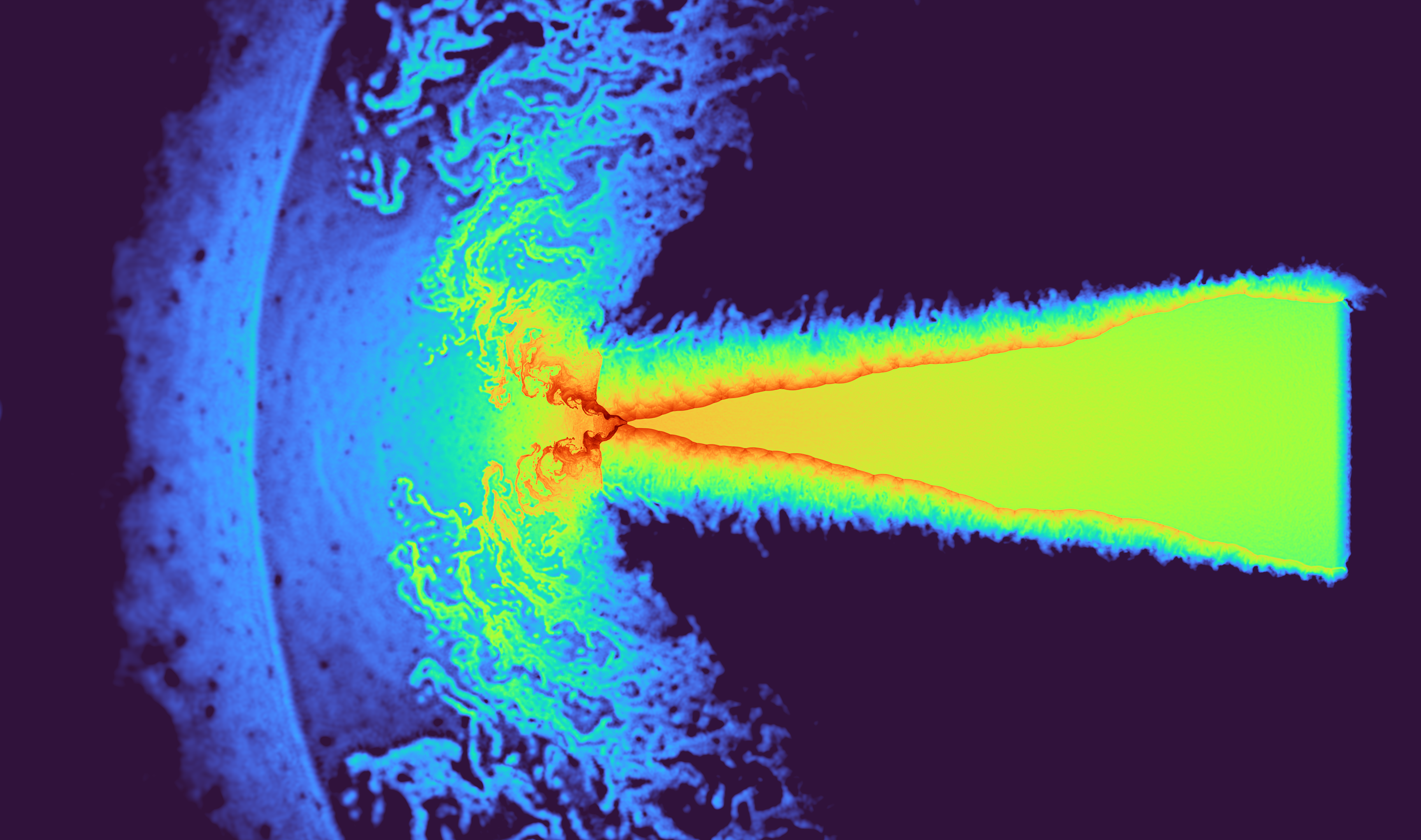}
         \caption{RN1-High: N = 2,000,000 particles}
         \label{fig:RN1High}
     \end{subfigure}
        \caption{Snapshots of the three variations on resolution of model RN1, taken at t = 49.35 h (Same time as last snapshot in Fig. \ref{fig:SnapshotsRecurrent}).}
        \label{fig:RN1comparison}
\end{figure}

The three resolution cases demonstrate excellent agreement in their global evolution, including the temporal development of disk disruption and shock front propagation (see figure \ref{fig:ComparisonRN1}). The primary macroscopic observables (shock velocities, mass ejection rates, and overall morphological evolution) remain consistent across all resolution levels. Excluding stochastic differences due to the different initial models, the principal resolution-dependent effect manifests in the characteristic scale of turbulent structures, with higher particle counts enabling the development of finer-scale features. This behavior is expected, as the effective viscosity scale decreases proportionally with the smoothing length reduction accompanying increased particle density (see figure \ref{fig:RN1comparison}).

It is important to recognize that turbulent flows exhibit a hierarchical cascade from large to small scales, extending to dissipation-dominated regimes that lie many orders of magnitude below current computational limits. Complete convergence in the traditional sense is therefore unattainable for such systems, as progressively smaller turbulent scales can always be resolved with sufficient computational resources. However, our results demonstrate convergence in the macroscopic level: the large-scale dynamics governing the interaction process remain invariant across the tested resolution range.

\end{document}